\documentclass{llncs}

\usepackage{latexsym}
\usepackage{algorithm,algorithmic}
\usepackage{graphicx}

\usepackage{txfonts}

\def\mO{\mbox{O}}
\newcommand{\nop}[1]{}

\title{A tight upper bound on the (2,1)-total labeling number of
outerplanar
graphs}
\author{Toru Hasunuma\inst{1}, Toshimasa Ishii\inst{2}, Hirotaka Ono\inst{3} and
 Yushi Uno\inst{4}}
\institute{
Department of Mathematical and Natural Sciences, The University of
Tokushima, Tokushima 770--8502
Japan. \email{hasunuma@ias.tokushima-u.ac.jp}
\and
Department of Information and Management Science, Otaru University of
Commerce, Otaru 047-8501, Japan. \email{ishii@res.otaru-uc.ac.jp}
\and
Department of Computer Science and Communication Engineering, Kyushu
University, Fukuoka 812-8581, Japan.
\email{ono@csce.kyushu-u.ac.jp}
\and
Department of Mathematics and Information Sciences, Graduate School of
Science, Osaka Prefecture University, Sakai 599-8531, Japan.
\email{uno@mi.s.osakafu-u.ac.jp}
}

\begin{document}
\maketitle

\begin{abstract}
A $(2,1)$-total labeling of a graph $G$ is an assignment $f$ 
from the vertex set $V(G)$ and the edge set $E(G)$ 
to the set $\{0,1,\ldots,k\}$ of nonnegative integers 
such that $|f(x)-f(y)|\ge 2$ if $x$ is a vertex and
$y$ is an edge incident to $x$,
and $|f(x)-f(y)|\ge 1$ if   $x$ and $y$ are a pair of  adjacent vertices
or a pair of 
adjacent
 edges, 
for all $x$ and $y$ in $V(G)\cup E(G)$. 
The $(2,1)$-total labeling number $\lambda^T_2(G)$ of a graph $G$
is defined as the minimum $k$
among all possible assignments.
In \cite{CW07},
Chen and Wang 
conjectured
that all   outerplanar graphs $G$ satisfy
  $\lambda^T_2(G) \leq \Delta(G)+2$, 
where  $\Delta(G)$ is the maximum degree of $G$,
while
they also  showed that
it is true 
for $G$ with 
$\Delta(G)\geq 5$. 
In this paper, we solve their conjecture
completely, by 
proving that 
 $\lambda^T_2(G) \leq \Delta(G)+2$ even in the case of $\Delta(G)\leq 4
$.
\end{abstract}

\section{Introduction}

In the channel/frequency assignment problems, 
we need to assign different
 frequencies to
`close' transmitters
so that 
 they 
can avoid interference. 
The
 $L(p,q)$-labelings of a graph  have been widely studied
as one of  important graph theoretical models of this problem.
An {\em $L(p,q)$-labeling} of a graph $G$ is an assignment $f$ 
from the vertex set $V(G)$ to the set $\{0,1,\ldots,k\}$
 of nonnegative integers 
such that $|f(x)-f(y)|\ge p$ if $x$ and $y$ are adjacent 
and $|f(x)-f(y)|\ge q$ if $x$ and $y$ are at distance 2, 
for all $x$ and $y$ in $V(G)$. 
We call this invariant (i.e.,  the minimum value $k$) 
the {\em $L(p,q)$-labeling number} and  denote it by $\lambda_{p,q}(G)$. 
Notice that we can use $k+1$ different labels when $\lambda_{p,q}(G)=k$ 
since we can use 0 as a label for conventional reasons. 
We can find many  results on $L(p,q)$-labelings 
in  comprehensive surveys by Calamoneri \cite{C06} and by Yeh \cite{Y06}. 

In \cite{WGM95}, Whittlesey et al. studied the 
$L(2,1)$-labeling number  
 of incidence graphs, where 
the {\em incidence graph} of a graph $G$ is the graph obtained from $G$
by replacing each edge $(v_i,v_j)$ with two edges $(v_i,v_{ij})$ and
$(v_{ij},v_j)$ introducing one new vertex $v_{ij}$.
Observe that an $L(p,1)$-labeling of the incidence graph of a given
graph $G$
can be regarded as 
an assignment $f$ 
from  $V(G) \cup E(G)$ 
to the set $\{0,1,\ldots,\ell\}$ of nonnegative integers 
such that $|f(x)-f(y)|\ge p$ if $x$ is a vertex and
$y$ is an edge incident to $x$,
and $|f(x)-f(y)|\ge 1$ if   $x$ and $y$ are a pair of  adjacent vertices
or a pair of 
adjacent
edges, 
for all $x$ and $y$ in $V(G)\cup E(G)$.
Such a labeling of $G$ is called
a {\em $(p,1)$-total labeling} of $G$,
  introduced by
 Havet and Yu  \cite{HY02,HY08}.
The {\em $(p,1)$-total labeling number} is defined as  
  the minimum value $\ell$, and  
denoted by $\lambda_p^T(G)$.

Notice that a  $(1,1)$-total labeling of $G$ is equivalent to a total coloring
of $G$.
Generalizing the Total Coloring Conjecture \cite{B65,V68}, 
Havet and Yu  \cite{HY02,HY08} conjectured that
\begin{equation}\label{conj:eq}
\lambda_p^T(G)\leq \Delta(G)+2p-1 
\end{equation} 
 holds for any graph $G$, where
$\Delta(G)$ denotes the maximum degree of a vertex in $G$.
They also showed
that  (i) $\lambda_p^T(G)\leq \min\{2\Delta(G)+p-1,\ \chi(G)+\chi'(G)+p-2\}$ 
for any graph
$G$ where $\chi(G)$ and $\chi'(G)$ denote the chromatic number
and
the chromatic index of $G$, respectively,
(ii) 
$\lambda_2^T(G)\leq 2\Delta(G)$
if $p=2$ and $\Delta(G)\geq 2$, 
 (iii) $\lambda_2^T(G)\leq 2\Delta(G)-1$ if $p=2$ and
$\Delta(G)$ is an odd $\geq 5$,
and (iv) $\lambda_p^T(G)\leq n+2p-2$
if $G$ is the complete graph where $n=|V(G)|$; 
 the conjecture (\ref{conj:eq}) is true if (a) $p\geq \Delta(G)$,
(b)
 $p=2$ and $\Delta(G)\leq 3$, or (c) $G$ is the complete graph.
By (i), it follows  that 
 $\lambda_p^T(G)\leq \Delta(G)+p$ for 
any bipartite graph  \cite{BMR07,HY02,HY08}
(by $\chi(G)\leq 2$ and $\chi'(G)=\Delta(G)$ \cite{K16}).
Also, Bazzaro et al. \cite{BMR07} investigated
that 
 $\lambda_p^T(G)\leq \Delta(G)+p+s$ for 
any $s$-degenerated graph 
(by $\chi(G)\leq s+1$ and $\chi'(G)\leq \Delta(G)+1$ \cite{V64}),
where an {\em $s$-degenerated graph} $G$ is a graph which
can be reduced to a trivial graph by successive removal of vertices 
with  degree at most $s$, and
that 
 $\lambda_p^T(G)\leq \Delta(G)+p+3$ for 
any planar graph 
(by the Four-Color Theorem).
\nop{
, and
that
 $\lambda_p^T(G)\leq \Delta(G)+p+1$ for 
any outerplanar graph other than an odd cycle 
(since any outerplaner graph is 2-degenerated, and 
any outerplaner graph other than an odd cycle satisfies $\chi'(G)= \Delta(G)$
 \cite{F75}).
}
They also 
\nop{showed that 
(i) $\lambda_p^T(G)\leq \Delta(G)+p$ for 
any bipartite graph,
(ii) $\lambda_p^T(G)\leq \Delta(G)+p+3$ for 
any planar graph, 
(iii) $\lambda_p^T(G)\leq \Delta(G)+p+1$ for 
any outerplanar graph,
and
(iv) $\lambda_p^T(G)\leq \Delta(G)+p+s$ for 
any $s$-degenerated graph.
They also} showed  sufficient conditions about 
$\Delta(G)$ and  girth for which the conjecture (\ref{conj:eq})
holds.
In \cite{MR06},
Montassier and Raspaud 
proved that
$\lambda_p^T(G) \leq \Delta(G)+2p-2$
when $p\geq 2$ and
$\Delta(G)$ and
the maximum average of $G$ degree satisfy some conditions.
In \cite{LLW09}, Lih et al. showed that
$\lambda_p^T(G) \leq \lfloor 3\Delta(G)/2 \rfloor+4p-3$ for 
any graph $G$.
They also investigated $\lambda_p^T(K_{m,n})$ of the complete bipartite
graphs $K_{m,n}$.

Let $G$ be an outerplanar graph.
In \cite{BMR07}, 
Bazzaro et al. pointed out that 
 $\lambda_p^T(G)\leq \Delta(G)+p+1$ for 
any outerplanar graph other than an odd cycle, 
since any outerplaner graph is 2-degenerated, and 
any outerplaner graph other than an odd cycle satisfies $\chi'(G)= \Delta(G)$
 \cite{F75}.
In particular,  Chen and Wang \cite{CW07}
conjectured that
in the case of  $p=2$, 
 $\lambda_2^T(G)\leq \Delta(G)+2$, which is a  better
bound than the conjecture (\ref{conj:eq}).
They also proved that this conjecture is true if
(i) $\Delta(G)\geq 5$,  (ii) $\Delta(G)=3$ and $G$ is
2-connected,
or (iii) $\Delta(G)=4$ and 
every two faces consisting of three vertices
have no  vertex in common.
In the case of $\Delta(G)=2$ (i.e., 
 $G$ is a path or a cycle),
we can easily see that $\lambda^T_2(G)\leq 4$,
since the incidence graph of a path (resp., a cycle) is also a path (resp., a
cycle),
and the $L(2,1)$-labeling number
 $\lambda_{2,1}(C_n)$ for a cycle $C_n$ with $n$ vertices
is at most 4 \cite{GY92}.
The cases of $\Delta(G)\in \{0,1\}$ are clear.
However, the general cases of $\Delta(G)\in\{3,4\}$ were left open.
In this paper, we solve Cheng and Wang's
conjecture completely, by showing the remaining cases of $\Delta(G)\in\{3,4\}$.
On the other hand, the bound $\Delta(G)+2$ is tight, since
there exist infinitely many outerplanar graphs $G$
such that $\lambda_2^T(G)=\Delta(G)+2$ if $\Delta(G)\geq 2$,
as investigated in   \cite{CW07}.


The rest of this paper is organized as follows. 
Section \ref{preliminaries-sec} gives basic definitions.
In Sections \ref{delta3-sec} and \ref{delta4-sec},
 we show that $\lambda_2^T(G)\leq \Delta(G)+2$
in the cases of $\Delta(G)=3$ and $\Delta(G)=4$,
recpectively.

\section{Preliminaries}\label{preliminaries-sec}

A graph $G$ is an ordered set of its vertex set $V(G)$ and edge set $E(G)$ 
and is denoted by $G=(V(G),E(G))$. 
We assume throughout this paper that a graph is undirected, 
simple and connected, unless otherwise stated. 
Therefore, an edge $e\in E(G)$ is an unordered pair of vertices $u$ and $v$, 
which are {\em end vertices} of $e$, 
and we often denote it by 
$e=(u,v)$.
For two graphs $G_1$ and $G_2$,
we denote $(V(G_1)\cup V(G_2), E(G_1)\cup E(G_2))$ by  $G_1+G_2$.
For a vertex set $V' \subseteq V(G)$, 
let $G-V'$  be the subgraph of $G$ induced by $V(G)-V'$ .
For an edge set $E'$, 
let $G-E'$ (resp., $G+E'$)  be the graph $(V,E-E')$ 
(resp., $(V\cup V(E'), E\cup E')$ where $V(E')$ is the set of
end vertices of edges in $E'$).
Two vertices $u$ and $v$ are {\em adjacent} if 
$(u,v)\in E(G)$.
Let $N_G(v)$ denote the set of neighbors of a vertex $v$ in $G$;
$N_G(v)=\{u \in V\mid (u,v) \in E(G)\}$.
The {\em degree} of a vertex $v$ is $|N_G(v)|$, 
and is denoted by $d_G(v)$. 
A vertex $v$ with $d_G(v)=k$ is called 
a {\em $k$-vertex}. 
We use $\Delta(G)$ (resp., $\delta(G)$) to denote the 
maximum (resp., minimum) degree of a graph $G$. 
A {\em path} $P$ in $G$ is a sequence $v_1,v_2,\ldots,v_{\ell}$ of 
vertices such that $(v_i,v_{i+1})\in E(G)$ for $i=1,2,\ldots,\ell-1$;
we may denote $P$ as $(v_1,v_2,\ldots,v_{\ell})$ and
let $V(P)=\{v_1,v_2,\ldots,v_{\ell}\}$
and $E(P)=\{(v_1,v_2),(v_2,v_3),\ldots,(v_{\ell-1},v_{\ell})\}$.
We often drop $G$ in these notations if there are no confusions.

The {\em vertex-connectivity} of a graph $G$, denoted by $\kappa(G)$,
is defined as the minimum cardinality of a vertex set $V' \subseteq V(G)$
such that $G-V'$ is disconnected or trivial.
A vertex $v$ is called a {\em cut vertex $($of $G)$} if
$G-v$ is disconnected.
Notice that a  graph $G$ has no cut vertex if and only if $\kappa(G)\geq 2$.
Two vertices $u$ and $u'$ are called {\em 2-connected} if they belong to the 
same connected component in $G-v$ for any cut vertex $v$ of $G$
with $v \notin \{u,u'\}$.
A subgraph $G'$ of $G$ is called a {\em 2-connected component}
if every two vertices in $G'$ are 2-connected and
$G'$ is maximal to this property.
It is not difficult to see that if $\kappa(G)=1$, then
there exists a 2-connected component which contains exactly one cut
vertex
of $G$.

A graph is called {\em planar}
 if it can be drawn
in the plane without generating a crossing by two edges.
For such a drawing of  $G$ in the plane, after
 omitting the line segments, the plane is divided into regions;
such a region is called a {\em face}.
A face whose vertex set is $\{u_1,u_2,\ldots,u_k\}$
with $(u_i,u_{i+1}) \in E(G)$, $i=1,2,\ldots,k$ (where $u_{k+1}=u_1$) 
is denoted by $[u_1u_2\cdots u_k]$.
We call a face consisting of
$k$ vertices  a {\em $k$-face}.

A planar graph $G$ is called {\em outerplanar} if it can be drawn
in the plane
 so that all vertices lie on the boundary of some face.
Such a drawing is referred to as an {\em outerplane graph}.
The face whose boundary contains all vertices of $G$
is called the {\em outer face} and all other faces are called
{\em inner faces}.
We call an edge belonging to the boundary of the outer face  an {\em outer edge}
and all other edges {\em inner edges}.
An inner face is called an {\em endface} if
its boundary contains exactly one inner edge (note that there exist at
least
two endfaces if $G$ contains an inner edge).

Let ${\cal L}_k=\{0,1,\ldots,k\}$ and
 $f: V(G)\cup E(G)\to {\cal L}_k$ 
be  a (2,1)-total labeling 
of $G$.
For an edge $e =(u,v) \in E(G)$,
we may denote $f(e)$ by $f(uv)$.
Let $\overline{f}$ denote the labeling such that 
$\overline{f}(z)=k-f(z)$ for each $z \in V(G) \cup E(G)$.
Notice that $\overline{f}$ is also a (2,1)-total labeling of $G$.

\section{Case $\Delta(G) =3$}\label{delta3-sec}

Let $G=(V,E)$ be an  outerplane  graph with $\Delta(G)=3$.
Chen and Wang \cite{CW07} showed that
if $\kappa(G)\geq 2$, then
$\lambda_2^T(G)\leq 5$.
\nop{where $|V|=n$ and
each vertex on the boundary of its outerface
is numbered
 $v_1,v_2,\ldots,v_n$
 in clockwise order.
}
First review their constructive proof of the case of $\kappa(G)\geq 2$.
Assume that 
on the boundary of its outerface,
each 3-vertex 
(i.e., each vertex which some inner edge is
incident to) 
is numbered
 $x_1,x_2,\ldots,x_p$
 in clockwise order.
Also assume that for $i=1,2,\ldots,p$,
each 2-vertex 
on the path  connecting $x_i$ and $x_{i+1}$ (on
the boundary of its outerface)
is numbered $y_1^i,y_2^i,\ldots,y_{q_i}^i$
in clockwise order, where $x_{p+1}=x_1$.
A  (2,1)-total labeling $f: V\cup E \to {\cal L}_5$
can be obtained by the following algorithm, which  slightly 
  generalizes Chen and Wang's method.

\begin{verse}
{\bf Algorithm LABEL-K2}

{\bf Input: }An outerplane graph $G$ with $\kappa(G)\geq 2$.

{\bf Output:} A total (2,1)-labeling  $f: V\cup E \to {\cal L}_5$ of $G$.

1. ~Assign  labels 0 and 1  alternately to a sequence 
$x_1, x_2, \ldots, x_p$  of 3-vertices.\\
2. ~For each $i=1,2,\ldots,p$,
label 2-vertices as follows. Without loss of generality,
let $f(x_i)=0$ and $f(x_{i+1})=1$.

~~2.1. If $q_i$ is even, then 
assign  labels 1 and 0 alternately to a sequence 
$y^i_1, y^i_2, \ldots, y^i_{q_i}$  of 2-vertices.

~~2.2. If $q_i$ is odd, then 
assign label 2 to  $y_j^i$ for some $j$,  and
 labels 1 and 0 alternately to a sequence 
$y^i_1, y^i_2, \ldots, y^i_{q_i}$  of 2-vertices other than $y_j^i$.

3. ~Assign label 3 to each inner edge.

4. ~Label  outer edges as follows.

~~4.1. ~If $|V(G)|$ is even, 	 
then assign  labels 4 and 5 alternately in clockwise order.

~~4.2. ~If $|V(G)|$ is odd, then 
pick up an arbitrary endface $F=[x_iy_1^i y_2^i \cdots y_{q_i}^i
 x_{i+1}]$.
Without loss of generality, let $f(x_i)=0$ and $f(x_{i+1})=1$.

~~~~4.2.1. ~If $q_i$ is an even or an odd with $q_i \geq 3$, then assign label 3 to $(y_1^i,y_2^i)$
and labels 4 and 5 alternately to all remaining outer edges.
Then if $f(y_1^i)=2$ or $f(y_2^i)=2$, 
then reassign labels for $y_1^i$, $y_2^i$, and $y_3^i$
as $f(y_1^i):=1$, $f(y_2^i):=0$, and $f(y_3^i):=2$.

\nop{~~~4.2.2. ~If $q_i$ is an odd with $q_i \geq 3$, then
assign label 3 to $(y_2^i,y_3^i)$
and labels 4 and 5 alternately to the remaining outer edges.
}

~~~4.2.2. ~If $q_i=1$, then
assign labels 5, 2, and 3 to
$y_1^i$, $(x_i,y_1^i)$, and $(y_1^i,x_{i+1})$, respectively,
and labels 4 and 5 alternately to all remaining outer edges.
Then if the label for two outer edges incident to
$x_i$ and $x_{i+1}$ other than $(x_i,y_1^i)$ and
$(y_1^i,x_{i+1})$ is 4 (resp., 5),
reassign a label for $(x_i,x_{i+1})$ as $f(x_i x_{i+1}):=5$ (resp., 4).
\end{verse}
\noindent
Observe  that for each inner edge $(x_i,x_j)$,
we have $f(x_i)\neq f(x_j)$, since otherwise
the subgraph $G'$ of $G$
 induced by $V(P_i)\cup V(P_{i+1}) \cup \cdots \cup V(P_{j-1})$ 
would have odd 3-vertices, a contradiction, where
$P_i$ denotes the path $(x_i, y_1^i,\ldots, y_{q_i}^i, x_{i+1})$.
Notice that the reassignment in Step 4.2.1 is necessary
because otherwise $|f(y_1^i)-f(y_1^i y_2^i))| \leq 1$ or
$|f(y_2^i)-f(y_1^i y_2^i))| \leq 1$ would hold.
Also notice that in Step 4.2.2, 
$f(e)=f(e')$ holds for 
two outer edges $e$ (resp.,
 $e'$) incident to
$x_i$ (resp., $x_{i+1}$) other than $(x_i,y_1^i)$
(resp., $(y_1^i,x_{i+1})$), since $|V(G)|$ is odd and $q_i=1$.
Thus,  a labeling $f$ obtained
from the above procedure is a (2,1)-total labeling of $G$.
Here, we have the following easy observation:

\vspace*{.2cm}

\noindent
{\bf Observation 1.} In Step 4.2,  we can start with assigning 
an arbitrary label in $\{4,5\}$ to an arbitrary outer 
edge for the label assignment
of outer edges except $(y_1^i,y_2^i)$ (resp., $(x_i,y_1^i)$ 
and $(y_1^i,x_{i+1})$) in Step 4.2.1 (resp., Step 4.2.2).

\vspace*{.2cm}

Moreover, we can see  the following property.

\begin{lemma}\label{extend1:lem}
Let $G=(V,E)$ be an  outerplane graph  with $\Delta(G)=3$,
 $G_1$ be a 2-connected component in $G$,
and $(u,v)$ be an inner edge in $G_1$.
Assume that $G_1-\{u,v\}$ contains
a connected component $G_2$ with no cut vertex in $G$
and $|V(G_2)|\geq 2$.
Let $u' \in V(G_2)$ {\rm (}resp., $v' \in V(G_2))$ be the
neighbor of $u$ {\rm (}resp., $v${\rm )} in $G$ 
 {\rm (}note that by $\Delta(G)=3$,
such $u'$ and $v'$ are uniquely determined and 
that by $|V(G_2)|\geq 2$, we have $u'\neq v'${\rm )}.
If $G'\triangleq (G-V(G_2))+\{(u,u'),(v,v')\}$ has a 
$($2,1$)$-total labeling  $f: V(G')\cup E(G') \to {\cal L}_5$
such that
\[
\begin{array}{l}
 \mbox{$f(u')\neq f(v')$ and 
{\rm (i)} $\{f(u'),f(v')\}\subseteq \{0,1\}$
and $\{f(uu'), f(vv')\} \subseteq \{4,5\}$ or}\\
\mbox{{\rm (ii)} $\{f(u'),f(v')\}\subseteq \{4,5\}$ and
$\{f(uu'), f(vv')\} \subseteq \{0,1\}$,}
\end{array}
\] then
there exists a $($2,1$)$-total labeling
  $g: V(G)\cup E(G) \to {\cal L}_5$.
\end{lemma}
\begin{proof}
By symmetry of labelings, we only consider the case
where $f(u')\neq f(v')$,
 $\{f(u'),f(v')\}\subseteq \{0,1\}$, 
and $\{f(uu'), f(vv')\} \subseteq \{4,5\}$.
Without loss of generality, let $f(u')=0$ and $f(v')=1$.
There are the two possible cases
(Case-1)  $f(uu')=f(vv')$
and (Case-2)  $f(uu') \neq f(vv')$.

(Case-1) ~Let $G_3$ be the graph obtained from $G_2$
by adding two new vertices $x$ and $y$
and three new edges $(u',x),(x,y)$, and $(y,v')$.
Notice that $G_3$ is an outerplane graph with $\kappa(G_3)\geq 2$.
There are the following two possible cases:
(1.1) both of
  $u'$ and $v'$ are  2-vertices or  3-vertices,
(1.2) otherwise.
 
(1.1) ~If $u'$ and $v'$ are both 2-vertices,
then add an inner edge $(u',v')$ to $G_3$ and
redefine $G_3$ by
the resulting graph.
By
applying Algorithm LABEL-K2 to $G_3$
as $x_1:=u'$, 
we can obtain a  $($2,1$)$-total labeling
 $f_1: V(G_3)\cup E(G_3) \to {\cal L}_5$
such that $f_1(u')=0$ and $f_1(v')=1$.
Moreover, by choosing an endface not
containing $\{x,y\}$ in Step 4.2 of Algorithm LABEL-K2, 
we can obtain  such a labeling $f_1$
that  $f_1(xu')=f_1(yv')=f(uu')~(=f(vv'))$ (note that
by Observation 1, we can start with assigning label $f(uu')$ to $(x, u')$
for the label assignment of outer edges). 
Observe that the labeling $g$ such that
$g(z)=f(z)$ for each $z \in V(G')\cup E(G')$
and
$g(z)=f_1(z)$  for each $z \in V \cup E -(V(G')\cup E(G'))$
is a  $($2,1$)$-total labeling of $G$.

(1.2) ~Assume that $u'$ is a 3-vertex and $v'$ is a 2-vertex
(the other case can be treated similarly).
Let 
 $f_1: V(G_3)\cup E(G_3) \to {\cal L}_5$
be  a  $($2,1$)$-total labeling
obtained 
by
applying Algorithm LABEL-K2 to $G_3$
as $x_1:=u'$;
 $f_1(u')=0$.
Moreover, as observed in (Case-1.1),
we can obtain such a $f_1$  that
  $f_1(xu')=f_1(yv')=f(uu')$.
Also, if $v'$ is not adjacent to a 3-vertex,
we can obtain such a $f_1$
that $f_1(v')=1$
by avoiding assigning label 2 to
any vertex in $\{x,y,v'\}$ in Step 2.2.
Observe that if $f_1(v')=1$, then
a  $($2,1$)$-total labeling $g$ of $G$
can be constructed by combining $f$ and $f_1$ similarly to
(Case-1.1).
Hence, we consider the case
where the vertex $w ~(\neq y)$  adjacent to $v'$ in $G_3$
is a 3-vertex; $f_1(w)=1$.
Then we can reassign 
labels for $x,y$, and $v'$
as any one of two possible labelings
(a) $f_1(x)=1, f_1(y)=0$, and $f_1(v')=2$
and 
(b) $f_1(x)=1, f_1(y)=2$, and $f_1(v')=0$
without violating the feasibility.
Observe that the labeling $g$ such that
$g(z)=f(z)$ for each $z \in V(G')\cup E(G')-\{v'\}$,
$g(z)=f_1(z)$  for each $z \in V \cup E -(V(G')\cup E(G'))$,
and $g(v')=\{0,2\}-\{f(v)\}$
is a  $($2,1$)$-total labeling of $G$.

(Case-2) ~Let $G_3$ be the graph obtained from $G_2$
by adding a new vertex $x$ 
and two new edges $(u',x)$ and $(x,v')$.
Notice that $G_3$ is an outerplane graph with $\kappa(G_3)\geq 2$.
There are the following two possible cases:
(2.1) both of
  $u'$ and $v'$ are  2-vertices or  3-vertices,
(2.2) otherwise.
 
(2.1) ~If $u'$ and $v'$ are both 2-vertices,
then add an inner edge $(u',v')$ to $G_3$ and
redefine $G_3$ by
the resulting graph.
By
applying Algorithm LABEL-K2 to $G_3$
as $x_1:=u'$, 
we can obtain a  $($2,1$)$-total labeling
 $f_1: V(G_3)\cup E(G_3) \to {\cal L}_5$
such that $f_1(u')=0$ and $f_1(v')=1$.
Moreover, by choosing an endface not 
containing $x$ in Step 4.2 of Algorithm LABEL-K2, 
we can obtain  such a labeling $f_1$
that  $f_1(xu')=f(uu')$ and $f_1(xv')=f(vv')$.
Observe that the labeling $g$ such that
$g(z)=f(z)$ for each $z \in V(G')\cup E(G')$
and
$g(z)=f_1(z)$  for each $z \in V \cup E -(V(G')\cup E(G'))$
is a  $($2,1$)$-total labeling of $G$.

(2.2) ~Assume that $u'$ is a 3-vertex and $v'$ is a 2-vertex
(the other case can be treated similarly).
Let 
 $f_1: V(G_3)\cup E(G_3) \to {\cal L}_5$
be  a  $($2,1$)$-total labeling
obtained 
by
applying Algorithm LABEL-K2 to $G_3$
as $x_1:=u'$;
 $f_1(u')=0$.
Moreover, as observed in (Case-2.1),
we can obtain such a $f_1$  that
  $f_1(xu')=f(uu')$ and $f_1(xv')=f(vv')$.
Also,  in the case where
 Step 2.2 is applied to the sequence of 
2-vertices containing $x$ and $v'$,
we can assign label 1 to $v'$ by
 assigning label 2 to $x$ (notice that
in this case, $v'$ is not adjacent to any 3-vertex because
the sequence of 2-vertices containing $v'$ consists of odd vertices). 
\nop{
we can obtain such a $f_1$
that $f_1(v')=1$
by avoiding assigning label 2 to
any vertex in $\{x,v'\}$ in Step 2.2.
}
Observe that if $f_1(v')=1$, then
a  $($2,1$)$-total labeling $g$ of $G$
can be constructed by combining $f$ and $f_1$ similarly to
(Case-2.1).
Hence, we consider the case
where  Step 2.1 is applied to the sequence of 
2-vertices containing $x$ and $v'$; $f_1(v')=0$.
Then we can reassign label 2 to $v'$ without violating the feasibility.
We can see that  the labeling $g$ such that
$g(z)=f(z)$ for each $z \in V(G')\cup E(G')-\{v'\}$,
$g(z)=f_1(z)$  for each $z \in V \cup E -(V(G')\cup E(G'))$,
and $g(v')=\{0,2\}-\{f(v)\}$
is a  $($2,1$)$-total labeling of $G$.
\qed
\end{proof}

~

By using this lemma, we can prove that Chen and Wang's conjecture is
true
in the case of $\Delta(G)=3$.

\begin{theorem}\label{delta3:theo}
If $G=(V,E)$ is an outerplane graph with $\Delta(G)=3$,
then  $\lambda_2^T(G)\leq 5$. 
\end{theorem}
\begin{proof}
 We prove this by induction on $k=|V(G)|+|E(G)|$.
The theorem clearly holds if $k \leq 7$.
Consider the case of $k \geq 8$ and assume that
for each $k' < k$, this theorem holds. 
We also assume that $G$ is connected, since otherwise
we can treat  each component separately.
Thus, $1 \leq \delta(G) \leq 2$.


Consider the case where $\delta(G)=1$.
Let $u_1$ be a vertex with $d(u_1)=1$.
By the induction hypothesis, $G-u_1$ has a (2,1)-total labeling $f: V(G-u_1)
 \cup E(G-u_1) \to {\cal L}_5$.
Let $u_2$ be the neighbor of $u_1$ in $G$ and
$u_3,u_4$ be vertices adjacent to $u_2$ in $G-u_1$
where $u_3=u_4$ may occur (note that $\Delta(G)=3$).
Hence we can extend $f$ to the edge $(u_1,u_2)$ and the vertex $u_1$ 
as follows:
 assign a label $a \in$
  ${\cal L}_5-$ $\{f(u_2)-1,f(u_2),f(u_2)+1,f(u_2 u_3), f(u_2 u_4) \}$
to  $(u_1,v_1)$, and then
a label in ${\cal L}_5-\{f(u_2),a-1,a,a+1\}$ to $u_1$.

Consider the case of $\delta(G)=2$.
In \cite{CW07}, Chen and Wang showed that
if $\kappa(G)\geq 2$, then 
$\lambda_2^T(G)\leq 5$. 
Hence, we here only consider  the case of $\kappa(G)=1$.
Let $G_1$ be a 2-connected component in $G$ which
has exactly one cut vertex $v_c$ of $G$.
By $\delta(G)=2$, we have $|V(G_1)|\geq 3$ and hence
$d_{G_1}(v_c)\geq 2$.
By $\kappa(G)=1$ and $\Delta(G)=3$,
it follows that
$d_{G_1}(v_c)=2$ holds and
$V(G_1)$ and $V-V(G_1)$ are connected by a bridge incident to $v_c$,
where a {\em bridge} is an edge whose deletion makes the graph
disconnected;
 denote this bridge by $(v_c,w)$ where $w \in V-V(G_1)$.
By the induction hypothesis, $H=(G-V(G_1))+\{(v_c,w)\}$ 
has a (2,1)-total labeling $f: V(H)
 \cup E(H) \to {\cal L}_5$.
By symmetry, it suffices to consider  the following three possible cases:
(Case-1) $f(v_c w)=5$,
(Case-2) $f(v_c w)=4$,
and
(Case-3) $f(v_c w)=3$.
In each case, we will extend $f$ to  
a (2,1)-total labeling $f': V \cup E \to {\cal L}_5$ of $G$.
Let $f'(x):=f(x)$ for every  $x \in V(H) \cup E(H)$.

First consider the case where $G_1$ has no inner edge.
Assume that on the boundary of its outerface in $G_1$,
each vertex 
is numbered
 $u_1 (=v_c),u_2,\ldots,u_r$
 in clockwise order.
Then, in each case of (Case-1)--(Case-3), 
we can obtain a (2,1)-total labeling $f'$ of $G$ in the following
 manner:

\begin{verse}
1. ~If $f'(v_c w)=3$, $r=3$, and $f'(w)\neq 0$, then 
let $f'(v_c)=0$, $f'(u_2):=1$, $f'(u_3):=5$,
$f'(v_c u_2):=5$, $f'(u_2 u_3):=3$, and $f'(u_3 v_c):=2$.
If $f'(v_c w)=3$, $r=3$, and $f'(w)= 0$, then 
let $f'(v_c)=5$, $f'(u_2):=2$, $f'(u_3):=0$,
$f'(v_c u_2):=0$, $f'(u_2 u_3):=5$, and $f'(u_3 v_c):=2$.

2. Unless $f'(v_c w)=3$ and $r=3$, then assign labels for vertices or
edges in $G_1$ as follows.

~~2.1. ~Assign a label in $\{0,1\}-\{f'(w)\}$ (say, 0) to $v_c$.

~~2.2. ~Assign  labels 1 and 0 alternately to 
$u_2, u_3, \ldots, u_r$. Then, if 
$r$ is odd, then reassign  label 2 to $u_r$.

~~2.3. ~If $f'(v_c w) \in \{3,4\}$ (resp., $f'(v_c w)=5$), then
assign  label  5 (resp., 4) to $(u_r, v_c)$,
and labels  4 and 5 (resp., 5 and 4) alternately to the sequence 
$(u_r,u_{r-1}),(u_{r-1},u_{r-2}),\ldots, (u_2,v_c)$
of outer edges in counter-clockwise order.
Then if $f'(v_c w)\in \{4,5\}$, then reassign 
label 3 to $(u_2,v_c)$, and
if $f'(v_c w)=3$ and $r$ is odd, then 
reassign label 3 to $(u_3,u_2)$  and
label 4 to $(u_2,v_c)$.
\end{verse}

Next consider the case where $G_1$ has an inner edge.
Assume that 
on the boundary of its outerface in $G_1$,
each 3-vertex 
is numbered
 $x_1,x_2,\ldots,x_p$
 in clockwise order.
Also assume that for $i=1,2,\ldots,p$,
each 2-vertex 
on the path  connecting $x_i$ and $x_{i+1}$ (on
the boundary of its outerface)
is numbered $y_1^i,y_2^i,\ldots,y_{q_i}^i$
in clockwise order, where $x_{p+1}=x_1$.
Also, if $q_i=0$, then let $y^i_{q_i}:=x_i$.
Let $P_i$ denote the path $(x_i, y_1^i,\ldots, y_{q_i}^i, x_{i+1})$.
Without loss of generality, assume that 
$v_c$ is a 2-vertex between $x_1$ and $x_2$ in $G_1$;
 $v_c=y_\ell^1$ for some $\ell$. 
Let $f_1: V(G_1)\cup E(G_1)\to {\cal L}_5$
be a (2,1)-total labeling obtained 
by applying Algorithm LABEL-K2 to $G_1$,
while 
(A) 
 choosing $y^1_j$ as a vertex other than  $v_c$
in Step 2.2 if  $q_1 \geq 3$, 
and
(B) choosing $y^1_j$ as a vertex not in $\{v_c\} \cup N_G(v_c)$
in Step 2.2
if (i)  $q_1 \geq 5$ or (ii) $q_1=3$ and $v_c \neq y_2^1$,
 and
(C) choosing an endface not containing $v_c$ in Step 4.2.

Consider the case where $v_c$ is adjacent to a 2-vertex in $G_1$;
$q_1 > 1$.
Without loss of generality, assume that $y_{\ell-1}^1$ is a 2-vertex.
Then by $q_1 > 1$ and
 the choice of $y_j^1$ in Step 2.2 of Algorithm LABEL-K2, 
we can see that
$f_1(v_c) \neq 2$; $f_1(v_c)\in \{0,1\}$.
Moreover, we can assume that $f_1(v_c) \in \{0,1\}-\{f'(w)\}$
since otherwise we recompute a (2,1)-total labeling $f'_1$ of $G_1$ 
 by starting with $x_2$ instead of $x_1$
in Step 1
of Algorithm LABEL-K2 and 
redefine $f_1$ by $f_1'$.
Also, we can assume that 
in the case of $f'(v_c w)=4$ (resp., 5),
if  $q_1 \neq 3$ or $v_c \neq y_2^1$,
then  $f_1(y_{\ell-1}^1 v_c)=4$ (resp., 5)
and
if  $q_1 =3$ and $v_c=y_2^1$,
then $f_1(v_c y')=4$ (resp., 5)
for $y' \in \{y_1^1,y_3^1\}$ with $f_1(y')\neq 2$
 (note that
this is possible by Observation 1 and that
in the case of 
 $q_1 =3$ and $v_c=y_2^1$,
exactly one of
$y_1^1$ and $y_3^1$ has label 2 in $f_1$). 
Let $f'(z):=f_1(z)$ for each $z \in V(G_1) \cup E(G_1)$.
Observe that if $f'(v_c w)=3$, then $f'$ is a (2,1)-total labeling of $G$.
In the case of  $f'(v_c w) \in \{4,5\}$,
if $q_1 \neq 3$ or  $v_c \neq y_2^1$  (resp., $q_1=3$
and  $v_c = y_2^1$),  reassign a label for $(y_{\ell-1}^1,
 v_c)$ (resp., $(y', v_c)$)
as  $f'(y_{\ell-1}^1 v_c):=3$ (resp., $f'(y' v_c):=3$).
Then, we can see that
\nop{
As a result, if $f'$ is still infeasible, then it follows 
by the choice of $y_j^1$ in Step 2.2
that  $q_1=3$, $y_2^1=v_c$, and
 $f_1(y_1^1)=2$; 
let   
$f_1(v_c)=0$  and
$f_1(y_1^3)=1$ (other cases can be treated similarly).
In this case, 
by reassigning   labels for $y_1^1$, $v_c$, and $y_1^3$ as
$f'(y_1^1):=0$, 
$f'(v_c):=1$ and $f'(y_1^3):=2$, 
we can obtain }
the resulting $f'$ is 
a (2,1)-total labeling  of $G$ since 
if $q_1\neq 3$ or  $v_c \neq y_2^1$, then
$f'(y_{\ell-1}^1)\neq 2$
 by the choice of
 $y_j^1$ in Step 2.2
and otherwise then
$f'(y')\neq 2$.

Finally, 
consider the case where $N_{G_1}(v_c)=\{x_1,x_2\}$; $q_1=1$.
Let $x_{p'}$ be a 3-vertex such that $(x_1,x_{p'})$ is an inner edge
of $G_1$.
Let $G_2$ be the subgraph of $G_1$ induced by
$V(P_1)\cup V(P_2) \cup \cdots \cup V(P_{p'-1})$. 
Let $u' \in V(G_1)-V(G_2)$ (resp., $v' \in V(G_1)-V(G_2)$) be the
 neighbor of $x_1$ (resp., $x_{p'}$).
Notice that the component in $G-\{x_1,x_{p'}\}$ containing $\{u',v'\}$
has no cut vertex in $G$.
Below,  we will show that 
 if $u'\neq v'$, then
 there exists a (2,1)-total labeling 
$g'$
of
 $(H+G_2)+\{(x_1,u'),(x_{p'},v')\}$ satisfying the conditions of 
Lemma~\ref{extend1:lem}.
Note that  in this case, 
Lemma~\ref{extend1:lem} ensures
the existence of a (2,1)-total labeling
$g: V \cup E \to {\cal L}_5$ of $G$.
Also, in the case of $u'=v'$,  we will show  directly  the existence of
 a (2,1)-total labeling
$g: V \cup E \to {\cal L}_5$ of $G$.

(Case-1)  $f(v_c w)=5$.

(1.1) Assume that $f(w)\neq 3$.

Let $f'(z):=\overline{f_1}(z)$ for each 
$z \in V(G_1)\cup E(G_1)$.
Then,  $f'$ is a (2,1)-total labeling of $G$
 (note that 
$f'(v_c)=3$).
\nop{
it follows that
one  of two vertices $u$ and $u'$ (say, $u$) adjacent to $v_c$ in $G_1$
 satisfies $f'(u)=\overline{f_1}(u)=3$.
This indicates that $q_1$ is odd and $u$ is a vertex chosen 
as $y_j^1$ in Step 2.2.
Hence, by reassigning a label for $u$ as
$f'(u):=\overline{f_1}(v_c)$, we can obtain a feasible labeling of $G$.
}

(1.2) Assume that $f(w)=3$. 

\nop{
Consider the case where $v_c$ is adjacent to a 2-vertex in $G_1$;
without loss of generality, $y_{\ell-1}^1$ is a 2-vertex.
Then we can assume that $f_1(y_{\ell-1}^1 v_c)=5$
since otherwise we can exchange  labels 4 and 5 in $f_1$
without violating its feasibility.
Let $f'(z):=f_1(z)$ for each $z \in V(G_1) \cup E(G_1)-\{(y_{\ell-1}^1,
 v_c)\}$ and $f'(y_{\ell-1}^1 v_c):=3$.
If $f'$ is infeasible, then it follows 
by the choice of $y_j^1$ in Step 2.2
that  $q_1=3$, $y_2^1=v_c$, and
 $f'(y_1^1)=2$ or 
  $f'(y_3^1)=2$;
let $f_1(y_1^1)=2$, $f_1(v_c)=0$  and
$f_1(y_1^3)=1$ (other cases can be treated similarly).
In this case, 
by reassigning   labels for $y_1^1$, $v_c$, and $y_1^3$ as
$f'(y_1^1):=0$, 
$f'(v_c):=1$ and $f'(y_1^3):=2$, 
we can obtain 
a (2,1)-total labeling of $G$.

Consider the case where $v_c$ is adjacent to both of $x_1$ and $x_2$.}

If $p'=2$ and $u'=v'$, then 
let $f'(x_1):=5$, $f'(v_c):=0$,  $f'(x_2):=2$, $f'(u'):=0$,
$f'(x_1 v_c):=3$, $f'(v_c x_2):=4$, $f'(x_1 x_2):=0$,
$f'(x_2 u'):=5$, and $f'(u'x_1):=2$.
Observe that the resulting labeling $f'$ is a (2,1)-total labeling of $G$.

If $p'=2$ and $u' \neq v'$, then
let $f'(x_1):=5$, $f'(v_c):=0$,  $f'(x_2):=4$, $f'(u'):=4$, $f'(v'):=5$,
$f'(x_1 v_c):=3$, $f'(v_c x_2):=2$, $f'(x_1 x_2):=0$,
$f'(x_2 v'):=1$, and $f'(u'x_1):=1$.
Observe that the resulting labeling $f'$ is feasible
for  $H+G_2+\{(x_1,u'),(x_2,v')\}$
and satisfies the conditions of Lemma~\ref{extend1:lem}.
\nop{
Notice that the component in $G-\{x_1,x_2\}$ containing $\{u',v'\}$
has no cut vertex in $G$.
Hence by Lemma~\ref{extend1:lem}, we can see that
$f'$ can be extended to
  a (2,1)-total labeling $g:V \cup E \to {\cal L}_5$ of $G$. 
}

Assume that $p'\neq 2$.
Then we can obtain a (2,1)-total labeling of $G$ in the following
 manner,
where ^^ ^^ assigning a label $k$ to $z \in V \cup E$'' means
that $f'(z):=k$.

\begin{verse}
1. Assign label 3 to each inner edge in $G_2$.

2. Let $f'(v_c):=0$. 
Assign labels 1 and 0 alternately to 3-vertices
 $x_2,x_3,\ldots,x_{p'-1}$ (note that as a result, $f'(x_{p'-1})=0$).
Assign labels to 2-vertices $y_j^i$ ($2 \leq i \leq p'-2$)  according to
Step 2 of algorithm LABEL-K2.  

3. Trace outer edges from $(v_c,x_2)$ to $(y^{p'-2}_{q_{p'-2}},x_{p'-1})$
on $\{(v_c,x_2)\}\cup E(P_2)\cup \cdots \cup E(P_{p'-2})$ in
 clockwise order and 
assign labels 4 and 5 alternately.

~~3.1. If $q_{p'-1}>0$, then assign labels to the remaining vertices or edges  as follows.
 Let $f'(x_1):=5$, $f'(x_{p'}):=4$,
$f'(x_1 v_c):=3$, $f'(x_1 x_{p'}):=2$, and $f'(x_{p'-1},y_1^{p'-1}):=2$.
Assign labels 0 and 1 alternately to outer edges
$(y_1^{p'-1},y_2^{p'-1}),(y_2^{p'-1},y_3^{p'-1}),\ldots, 
(y_{q_{p'-1}}^{p'-1},x_{p'}),$
$(x_{p'},v')$. 
Moreover, if $q_{p'-1}$ is even (resp., odd), then
assign labels 4 and 5 (resp., 5 and 4) alternately to 2-vertices
$y_1^{p'-1},y_2^{p'-1},\ldots, y_{q_{p'-1}}^{p'-1}$.
If $u'=v'$, then let $f'(u'):=3$ and $f'(u' x_1
 ):=\{0,1\}-\{f'(x_{p'}u')\}$.
If $u'\neq v'$,  then let $f'(u'):=4$, $f'(v'):=5$, and $f'(u' x_1):=1$,
and extend $f'$ to a (2,1)-total labeling  of $G$ according to Lemma~\ref{extend1:lem}.

~~3.2. If $q_{p'-1}=0$ and $u'\neq v'$, then 
let $f'(x_1):=f'(v'):=4$, $f'(x_{p'}):=f'(u'):=5$,
$f'(x_1x_{p'}):=0$,
$f(x_{p'-1}x_{p'}):=f(x_1 v_c):=2$, and
$f'(x_1 u'):=f'(x_{p'}v'):=1$, and
extend
 $f'$ to a (2,1)-total labeling  of $G$ according to
 Lemma~\ref{extend1:lem}.

~~3.3. If $q_{p'-1}=0$, $u'= v'$ and $f'(y_{q_{p'-2}}^{p'-2}x_{p'-1})=4$,
 then 
let $f'(x_1):=5$, $f'(x_{p'}):=3$, $f'(u'):=4$,
$f'(x_1x_{p'}):=1$,
$f(x_{p'-1}x_{p'}):=5$, $f(x_1 v_c):=3$, and
$f'(x_1 u'):=2$, and $f'(x_{p'}u'):=0$.

~~3.4. If $q_{p'-1}=0$, $u'= v'$ and  $f'(y_{q_{p'-2}}^{p'-2}x_{p'-1})=5$,
 then 
let $f'(x_1):=5$, $f'(x_{p'}):=2$, $f'(u'):=0$,
$f'(x_1x_{p'}):=0$,
$f(x_{p'-1}x_{p'}):=4$, $f(x_1 v_c):=3$, and
$f'(x_1 u'):=2$, and $f'(x_{p'}u'):=5$.
\end{verse}

(Case-2)  $f(v_c w)=4$.

(2.1) Assume that $f(w)\neq 0$.

If  $p'=2$ and $u'=v'$, 
\nop{Let $u'$ (resp., $v'$) be the neighbor of $x_1$ (resp., $x_{p'}$) 
which is neither  $v_c$ nor $x_2$ (resp., $x_1$). }
 then
let $f'(x_1):=5$, $f'(v_c):=0$,  $f'(x_2):=2$, $f'(u')=0$,
$f'(x_1 v_c):=3$, $f'(v_c x_2):=5$, $f'(x_1 x_2):=0$,
$f'(x_2 u'):=4$, and $f'(u'x_1):=2$.

If $p'=2$ and  $u' \neq v'$, then
let $f'(x_1):=f'(v'):=5$, $f'(v_c):=0$,  $f'(x_2):=f'(u')=:4$, 
$f'(x_1 v_c):=3$, $f'(v_c x_2):=2$, $f'(x_1 x_2):=0$,
$f'(x_2 v'):=1$,  $f'(u'x_1):=1$.
Observe that the resulting labeling $f'$ is feasible
for  $H+G_2+\{(x_1,u'),(x_2,v')\}$
and satisfies the conditions of Lemma~\ref{extend1:lem}.
%

The  case of $p'\neq 2$ can be treated similarly to (Case-1.2),
by starting with assigning to $(v_c,x_2)$ label 5 instead of
 label 4.

(2.2) Assume that $f(w)=0$. 

\nop{
Consider the case where $v_c$ is adjacent to a 2-vertex in $G_1$;
without loss of generality, $y_{\ell-1}^1$ is a 2-vertex.
Then we can assume that 
 $f_1(y_{\ell-1}^1 v_c)=4$,
since otherwise we can exchange  labels 4 and 5 in $f_1$
without violating its feasibility.
Let $f'(z):=f_1(z)$ for each $z \in V(G_1) \cup E(G_1)-\{(y_{\ell-1}^1,
 v_c)\}$  and $f'(y_{\ell-1}^1 v_c):=3$.
If $f'$ is infeasible, then it follows 
by the choice of $y_j^1$ in Step 2.2
that  $q_1=3$, $y_2^1=v_c$, and
 $f'(y_1^1)=2$ or  $f'(y_3^1)=2$; 
let   $f_1(y_1^1)=2$, 
$f_1(v_c)=0$  and
$f_1(y_1^3)=1$ (other cases can be treated similarly).
In this case, 
by reassigning   labels for $y_1^1$, $v_c$, and $y_1^3$ as
$f'(y_1^1):=0$, 
$f'(v_c):=1$ and $f'(y_1^3):=2$, 
we can obtain 
a (2,1)-total labeling of $G$.

Consider the case where $N_{G_1}(v_c)=\{x_1,x_2\}$.
Let $x_{p'}$ be a 3-vertex such that $(x_1,x_{p'})$ is an inner edge
of $G_1$.
Let $G_2$ be the subgraph of $G_1$ induced by
$V(P_1)\cup V(P_2) \cup \cdots \cup V(P_{p'-1})$. 
Let $u' \in V(G_1)-V(G_2)$ (resp., $v' \in V(G_1)-V(G_2)$) be the neighbor of $x_1$ (resp., $x_{p'}$).
}

If $p'=2$ and $u'=v'$, then 
let $f'(v_c):=1$,
 $f'(x_1):=5$, $f'(x_2):=3$, $f'(u'):=4$,
$f'(x_1 v_c):=3$,
$f'(v_c x_2):=5$, 
$f'(x_1 x_2):=0$,  $f'(x_1 u'):=2$,
and  $f'(x_2u'):=1$.

If $p'=2$ and $u' \neq v'$, then
$f'(v_c):=1$,
 $f'(x_1):=5$, $f'(x_2):=3$, $f'(u'):=4$,
$f'(v'):=5$,
$f'(x_1 v_c):=3$,
$f'(v_c x_2):=5$, 
$f'(x_1 x_2):=0$,  $f'(x_1 u'):=1$,
and  $f'(x_2v'):=1$.
Observe that the resulting labeling is feasible
for  $H+G_2+\{(x_1,u'),(x_2,v')\}$
and satisfies the conditions of Lemma~\ref{extend1:lem}.
\nop{
 Notice that the component in $G-\{x_1,x_2\}$ containing $\{u',v'\}$
has no cut vertex in $G$.
Hence by Lemma~\ref{extend1:lem}, we can see that
$f'$ can be extended to
  a (2,1)-total labeling $g: V \cup E \to {\cal L}_5$ of $G$. 
}

Assume that $p'\neq 2$.
Then we can obtain a (2,1)-total labeling of $G$ in the following manner.

\begin{verse}
1. Assign label 3 to each inner edge in $G_2$.

\nop{
2. Let $f'(v_c):=0$. 
Assign labels 1 and 0 alternately to 3-vertices
 $x_2,x_3,\ldots,x_{p'-1}$.
Assign labels to 2-vertices $y_j^i$ ($2 \leq i \leq p'-2$)   according to
Step 2 of algorithm LABEL-K2.  
}

2. Trace outer edges from $(v_c,x_2)$ to $(y^{p'-1}_{q_{p'-1}},x_{p'})$
on $\{(v_c,x_2)\}\cup E(P_2)\cup \cdots \cup E(P_{p'-1})$ in
 clockwise order 
and
assign labels 5 and 4 alternately.

~~2.1. If $f'(y^{p'-1}_{q_{p'-1}}x_{p'})=4$  and
 $u' \neq v'$, 
then assign labels for the remaining  vertices and edges as follows.
Let $f'(v_c):=2$. 
Assign labels 1 and 0 alternately to 3-vertices
 $x_2,x_3,\ldots,x_{p'-1}$ (notice that $f'(x_{p'-1})=0$).
Assign labels to 2-vertices $y_j^i$ ($2 \leq i \leq p'-2$)  according to
Step 2 of algorithm LABEL-K2.  
 Let $f'(x_1):=3$, $f'(x_{p'}):=2$, $f'(u'):=4$,
$f'(v'):=5$,
$f'(x_1 v_c):=0$, $f'(x_1 x_{p'}):=5$,  $f'(x_1 u'):=1$,
and  $f'(x_{p'}v'):=0$.
Assign labels 1 and 0  alternately to 2-vertices
$y_1^{p'-1},y_2^{p'-1},\ldots, y_{q_{p'-1}}^{p'-1}$.
Then, extend $f'$ to a (2,1)-total labeling  of $G$ 
according to Lemma~\ref{extend1:lem}.

~~2.2. If $f'(y^{p'-1}_{q_{p'-1}}x_{p'})=5$  and
 $u' \neq v'$, 
then assign labels for  the remaining  vertices and edges as follows.
Let $f'(v_c):=1$. 
Assign labels 0 and 1 alternately to 3-vertices
 $x_2,x_3,\ldots,x_{p'-1}$ (notice that $f'(x_{p'-1})=1$).
Assign labels to 2-vertices $y_j^i$ ($2 \leq i \leq p'-2$)  according to
Step 2 of algorithm LABEL-K2.  
 Let $f'(x_1):=5$, $f'(x_{p'}):=3$, $f'(u'):=4$,
$f'(v'):=5$,
$f'(x_1 v_c):=3$, $f'(x_1 x_{p'}):=0$,  $f'(x_1 u'):=1$,
and  $f'(x_{p'}v'):=1$.
Assign labels 0 and 1  alternately to 2-vertices
$y_1^{p'-1},y_2^{p'-1},\ldots, y_{q_{p'-1}}^{p'-1}$.
Then, extend $f'$ to a (2,1)-total labeling  of $G$ 
according to Lemma~\ref{extend1:lem}.

~~2.3. If $f'(y^{p'-1}_{q_{p'-1}}x_{p'})=4$  and
 $u' = v'$,
then assign labels for the remaining  vertices and edges as follows.
Let $f'(v_c):=2$. 
Assign labels 1 and 0 alternately to 3-vertices
 $x_2,x_3,\ldots,x_{p'-1}$ (notice that $f'(x_{p'-1})=0$).
Assign labels to 2-vertices $y_j^i$ ($2 \leq i \leq p'-2$)  according to
Step 2 of algorithm LABEL-K2.  
 Let $f'(x_1):=3$, $f'(x_{p'}):=2$, $f'(u'):=5$,
$f'(x_1 v_c):=0$, $f'(x_1 x_{p'}):=5$,  $f'(x_1 u'):=1$,
and  $f'(x_{p'}u'):=0$.
Assign labels 1 and 0  alternately to 2-vertices
$y_1^{p'-1},y_2^{p'-1},\ldots, y_{q_{p'-1}}^{p'-1}$.

~~2.4. If $f'(y^{p'-1}_{q_{p'-1}}x_{p'})=5$  and
 $u' = v'$,
then assign labels for  the remaining  vertices and edges as follows.
Let $f'(v_c):=1$. 
Assign labels 0 and 1 alternately to 3-vertices
 $x_2,x_3,\ldots,x_{p'-1}$ (notice that $f'(x_{p'-1})=1$).
Assign labels to 2-vertices $y_j^i$ ($2 \leq i \leq p'-2$)  according to
Step 2 of algorithm LABEL-K2.  
 Let $f'(x_1):=5$, $f'(x_{p'}):=3$, $f'(u'):=4$,
$f'(x_1 v_c):=3$, $f'(x_1 x_{p'}):=0$,  $f'(x_1 u'):=2$,
and  $f'(x_{p'}u'):=1$.
Assign labels 0 and 1  alternately to 2-vertices
$y_1^{p'-1},y_2^{p'-1},\ldots, y_{q_{p'-1}}^{p'-1}$.
\end{verse}

(Case-3)  $f(v_c w)=3$.

\nop{
(3.1) Assume that $q_1\neq 1$.

Consider the case where $f(w)=0$.
Notice that $f_1(v_c)\neq 2$ by the choice of $y_1^j$ in Step 2.2 of
Algorithm LABEL-K2.
Hence,  $f_1(v_c) \in \{0,1\}$.
If $f_1(v_c)=0$,
then recompute a (2,1)-total labeling $f'_1$ of $G$ 
 by starting with $x_2$ instead of $x_1$
in Step 1
of Algorithm LABEL-K2; $f'_1(v_c)=1$ and 
redefine $f_1$ by $f_1'$.
Let $f'(z):=f_1(z)$ for each 
$z \in V(G_1)\cup E(G_1)$.
Observe that $f'$ is a (2,1)-total labeling of $G$.

Consider the case where $f(w)\neq 0$.
Notice that $f_1(v_c)\neq 2$ by the choice of $y_1^j$.
Hence,  $f_1(v_c) \in \{0,1\}$.
If $f_1(v_c)=1$,
then recompute a (2,1)-total labeling $f'_1$ of $G$ 
 by starting with $x_2$ instead of $x_1$
in Step 1
of Algorithm LABEL-K2; $f'_1(v_c)=0$ and 
redefine $f_1$ by $f_1'$.
Let $f'(z):=f_1(z)$ for each 
$z \in V(G_1)\cup E(G_1)$.
Observe that $f'$ is a (2,1)-total labeling of $G$.

(3.2) Assume that $q_1 = 1$.

By $q_1=1$,
we have $N_{G_1}(v_c)=\{x_1,x_2\}$.
Let $x_{p'}$ be a 3-vertex such that $(x_1,x_{p'})$ is an inner edge
of $G_1$.
Let $G_2$ be the subgraph of $G_1$ induced by
$V(P_1)\cup V(P_2) \cup  \cdots \cup V(P_{p'-1})$. 
Let $u' \in V(G_1)-V(G_2)$ (resp., $v' \in V(G_1)-V(G_2)$) be the neighbor of $x_1$ (resp., $x_{p'}$).
}

First consider the case where $p'=2$ and $u'=v'$.
If $f(w)\neq 5$, then
let $f'(x_1):=0$, $f'(v_c):=5$,  $f'(x_2):=2$, $f'(u'):=1$,
$f'(x_1 v_c):=2$, $f'(v_c x_2):=0$, $f'(x_1 x_2):=5$,
$f'(x_2 u'):=4$, and $f'(u'x_1):=3$.
If $f( w)= 5$, then
let $f'(x_1):=4$, $f'(v_c):=0$,  $f'(x_2):=2$, $f'(u'):=3$,
$f'(x_1 v_c):=2$, $f'(v_c x_2):=4$, $f'(x_1 x_2):=0$,
$f'(x_2 u'):=5$, and $f'(u'x_1):=1$.

Next consider the case where $p'=2$ and $u' \neq v'$.
If  $f( w)\neq 5$, then
let $f'(x_1):=0$, $f'(v_c):=5$,  $f'(x_2):=2$, $f'(u'):=1$, $f'(v'):=0$,
$f'(x_1 v_c):=2$, $f'(v_c x_2):=0$, $f'(x_1 x_2):=5$,
$f'(x_2 v'):=4$, and $f'(u'x_1):=4$.
If  $f( w)= 5$, then
let $f'(x_1):=3$, $f'(v_c):=0$,  $f'(x_2):=5$, $f'(u'):=5$, $f'(v'):=4$,
$f'(x_1 v_c):=5$, $f'(v_c x_2):=2$, $f'(x_1 x_2):=0$,
$f'(x_2 v'):=1$, and $f'(u'x_1):=1$.
Observe that in both cases, the resulting labeling is feasible
for  $H+G_2+\{(x_1,u'),(x_{p'},v')\}$
and satisfies the conditions of Lemma~\ref{extend1:lem}.

\nop{
Then by Lemma~\ref{extend1:lem}, 
$f'$ can be extended to
  a (2,1)-total labeling $g:V \cup E \to {\cal L}_5$ of $G$. 
}

Finally consider the case where $p'\neq 2$.
We divide this case into the following two subcases:
(3.1) $f'( w)\neq 0$ and
(3.2) $f'( w)=0$.

(3.1) Assume that $f'( w)\neq 0$.

We can obtain a (2,1)-total labeling of $G$ in the following manner.

\begin{verse}
1. Assign label 3 to each inner edge in $G_2$.

2. Let $f'(v_c):=0$. 
Assign labels 1 and 0 alternately to 3-vertices
 $x_2,x_3,\ldots,x_{p'-1}$ (note that as a result, $f'(x_{p'-1})=0$).
Assign labels to 2-vertices $y_j^i$ ($2 \leq i \leq p'-2$)  according to
Step 2 of algorithm LABEL-K2.  
Assign labels 1 and 0  alternately to 2-vertices
$y_1^{p'-1},y_2^{p'-1},\ldots, y_{q_{p'-1}}^{p'-1}$.

3. Trace outer edges from $(v_c,x_2)$ to $(y^{p'-1}_{q_{p'-1}},x_{p'})$
on $\{(v_c,x_2)\}\cup E(P_2)\cup \cdots \cup E(P_{p'-1})$ in
 clockwise order 
 and 
assign labels 4 and 5 (resp., 5 and 4) alternately 
if the number of outer edges in $G_2$
 is
even (resp., odd).
Notice that as a result, the edge $(y^{p'-1}_{q_{p'-1}},x_{p'})$
 has label 5.

~~3.1. If  $u' \neq v'$,
then let
  $f'(x_1):=5$, $f'(x_{p'}):=3$, $f'(u'):=4$,
$f'(v'):=5$,
$f'(x_1 v_c):=2$, $f'(x_1 x_{p'}):=0$,  $f'(x_1 u'):=1$,
and  $f'(x_{p'}v'):=1$.
Then, extend $f'$ to a (2,1)-total labeling  of $G$ 
according to Lemma~\ref{extend1:lem}.

~~3.2. If  $u' = v'$, 
then 
let
  $f'(x_1):=5$, $f'(x_{p'}):=2$, $f'(u'):=0$,
$f'(x_1 v_c):=2$, $f'(x_1 x_{p'}):=0$,  $f'(x_1 u'):=3$,
and  $f'(x_{p'}u'):=4$.
\end{verse}

(3.2) Assume that $f'( w)= 0$.

We can obtain a (2,1)-total labeling of $G$ in the following manner.

\begin{verse}
1. Assign label 2 to each inner edge in $G_2$.

2. Let $f'(v_c):=5$. 
Assign labels 4 and 5 alternately to 3-vertices
 $x_2,x_3,\ldots,x_{p'-1}$  (note that as a result, $f'(x_{p'-1})=5$).
Assign labels to 2-vertices $y_j^i$ ($2 \leq i \leq p'-2$) 
 according to
Step 2 of Algorithm LABEL-K2',
where Algorithm LABEL-K2' denotes the algorithm
obtained from Algorithm LABEL-K2 by replacing each label $i$
to be assigned with $5-i$.
\nop{
 $5-f_2(y_j^i)$, where
$f_2(y_j^i)$ denotes a label assigned to $y_j^i$  
.}
Assign labels 4 and 5  alternately to 2-vertices
$y_1^{p'-1},y_2^{p'-1},\ldots, y_{q_{p'-1}}^{p'-1}$.

3. Trace outer edges from $(v_c,x_2)$ to $(y^{p'-1}_{q_{p'-1}},x_{p'})$
on $\{(v_c,x_2)\}\cup E(P_2)\cup \cdots \cup E(P_{p'-1})$ in
 clockwise order 
 and 
assign labels 1 and 0 (resp., 0 and 1) alternately 
if the number of outer edges in $G_2$
 is
even (resp., odd).
Notice that as a result, the edge $(y^{p'-1}_{q_{p'-1}},x_{p'})$
 has label 0.

~~3.1. If  $u' \neq v'$,
then let
  $f'(x_1):=0$, $f'(x_{p'}):=2$, $f'(u'):=1$,
$f'(v'):=0$,
$f'(x_1 v_c):=2$, $f'(x_1 x_{p'}):=4$,  $f'(x_1 u'):=5$,
and  $f'(x_{p'}v'):=5$.
Then, extend $f'$ to a (2,1)-total labeling  of $G$ 
according to Lemma~\ref{extend1:lem}.

~~3.2. If  $u' = v'$, 
then 
let
  $f'(x_1):=0$, $f'(x_{p'}):=2$, $f'(u'):=1$,
$f'(x_1 v_c):=2$, $f'(x_1 x_{p'}):=4$,  $f'(x_1 u'):=3$,
and  $f'(x_{p'}u'):=5$.
\end{verse}
\qed
\end{proof}

\section{Case $\Delta(G) =4$}\label{delta4-sec}

Let $G=(V,E)$ be an  outerplane  graph.
The following structural properties of outerplane graphs
are known.

\nop{where $|V|=n$ and
each vertex on the boundary of its outerface
is numbered
 $v_1,v_2,\ldots,v_n$
 in clockwise order.
}
\nop{
For an inner edge $(v_i,v_j) \in E$, 
let $G[v_i,v_j]$ denote the subgraph of $G$ induced by
$\{v_i,v_{i+1},\ldots,v_{j-1},v_j\}$.
}


\begin{theorem}\label{config:theo}
{\rm \cite{WZ99}}
Every outerplane graph $G$ with $\delta(G)=2$ contains one of the
 following
configurations:\\
{\rm (C1)} two adjacent 2-vertices $u_1$ and $u_2$.\\
{\rm (C2)} a 3-face $[u_1 u_2 u_3]$ with $d(u_1)=2$ and $d(u_2)=3$. \\
{\rm (C3)} two 3-faces $[u_1 u_2 u_3]$ and $[u_3 u_4 u_5]$ such that
$d(u_3)=4$ and $d(u_2)=d(u_4)=2$.
\end{theorem}

\noindent
In \cite{CW07}, Chen and Wang proved that
if $G$ does not contain (C3) and $\Delta(G)=4$ holds, then  
$\lambda_2^T(G)\leq 6$.
Their proof utilizes the property that
if $G$ contains (C1) or (C2), then
a (2,1)-total labeling of $G$ can be extended from
a (2,1)-total labeling of some proper subgraph of $G$.
Here, we investigate  the case where 
$G$  contains neither (C1) nor (C2),
and  derive a new structural property, i.e, if
$G$  contains neither (C1) nor (C2) and
$\Delta(G)=4$ holds,
then $G$ contains a new configuration as shown in Lemma~\ref{config2:lem}
below,
which includes (C3) as a subconfiguration.

\begin{figure*}[ht]
  \begin{center}
   \resizebox{4truein}{3.2truein}{\includegraphics{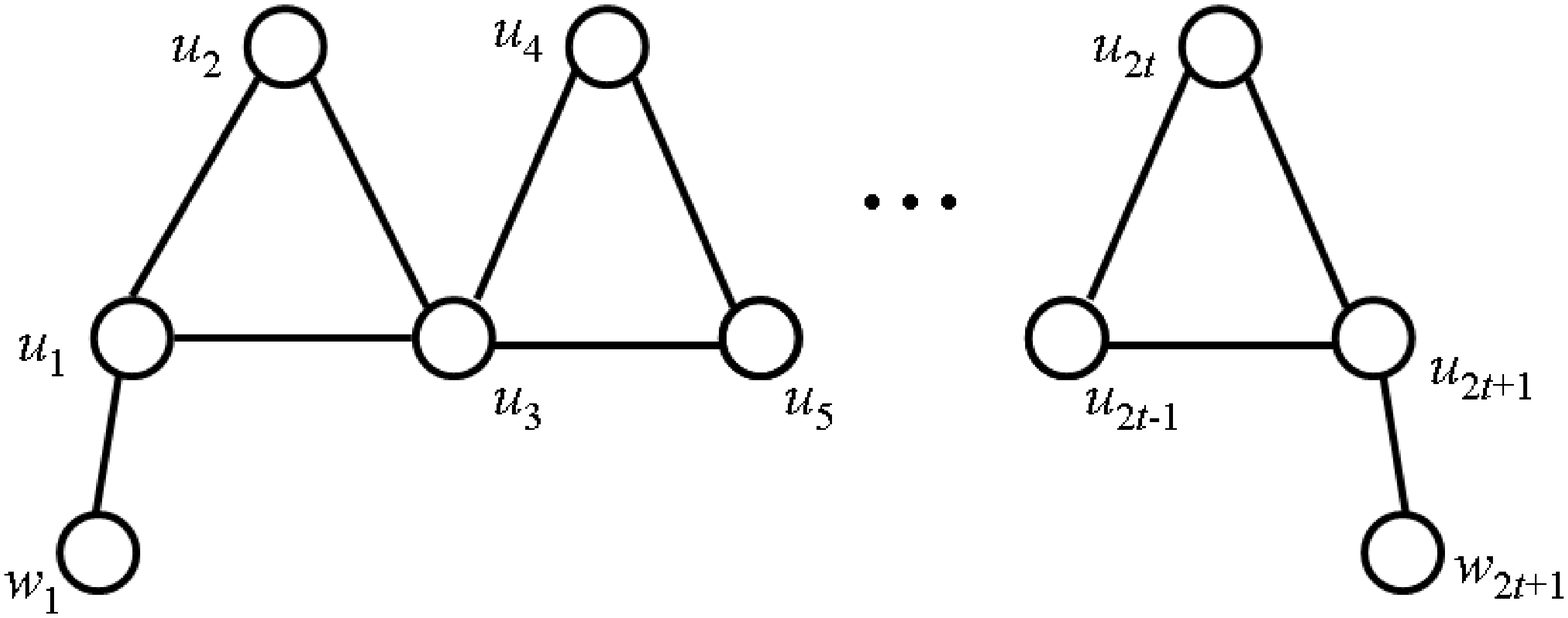}}
  \end{center}
\vspace*{-2.4cm}
 \caption{A chain of 3-faces.}
 \label{fig:chain}
\end{figure*}

Assume that $\Delta(G)=4$.
Define 
a {\em chain of 3-faces}
as   a sequence 
\[
{\cal C}=\{ [u_1 u_2 u_3],[u_3 u_4 u_5],\ldots,[u_{2t-1} u_{2t} u_{2t+1}]\} ~(t\geq 2)
\]
of 3-faces 
such that $(u_{2i-1},u_{2i})$ and
$(u_{2i},u_{2i+1})$ are on the outer face of $G$ and
$(u_{2i-1},u_{2i+1})$ is an inner edge for each $i=1,2,\ldots,t$
(see Fig. \ref{fig:chain}).
Notice that $d(u_{2i})=2$ holds for each $i=1,2,\ldots,t$
and that
 $d(u_{2i+1})=4$ holds for each $i=1,2,\ldots,t-1$ by $\Delta(G)=4$;
no vertex  other than $u_{2i-1},u_{2i}, u_{2i+2}$, and
$u_{2i+3}$ is adjacent to $u_{2i+1}$.

\begin{lemma}\label{config2:lem}
If $G$ is an outerphane graph
with $\Delta(G)=4$ and
 $\delta(G)=2$ which  contains neither {\rm (C1)}
nor {\rm (C2)}, then $G$ has a chain
 $\{[u_1 u_2 u_3],[u_3 u_4 u_5],\ldots,[u_{2t-1} u_{2t} u_{2t+1}]\}$
 of 3-faces such that $(u_1,u_{2t+1})$ is an inner edge.
\end{lemma}
\begin{proof}
Let $G$ be an outerphane graph
with $\delta(G)=2$ which  contains neither {\rm (C1)}
nor {\rm (C2)}.
Let $G'$ be a 2-connected component of $G$ which 
contains exactly one cut vertex $v_c$ of $G$ if $G$ is not 2-connected,
$G'=G$ otherwise.
Notice that $|V(G')|\geq 3$ by $\delta(G)=2$.
Then $G'$ has an inner edge since otherwise
$G$ contains (C1). 
There are the following two possible cases: 
(Case-I) every inner edge in $G'$ belongs to some endface,
(Case-II) otherwise.
Assume that each vertex on the boundary of the outerface of $G'$
is numbered
 $v'_1,v'_2,\ldots,v'_t$
 in clockwise order.
Notice that each endface not containing $v_c$
as a 2-vertex in $G'$ is a 3-face since otherwise it
would contain (C1). 

(Case-I)
Notice that there are at least two end faces in $G'$.
Without loss of generality, 
let $[v_1'v_2'v_3']$ be an endface of $G'$ with 
 $v_c \notin \{v_2',v_3'\}$; 
 $d_G(v_2')=d_{G'}(v_2')=2$. 
Then 
 an inner edge other than $(v_1',v_3')$ is incident to  $v_3'$, 
 since otherwise  $d_G(v_3')=3$ holds and  the 3-face $[v_1'v_2'v_3']$
would satisfy the conditions of (C2).  
By the assumption of Case-I, the inner edge belongs to some endface,
and hence we can see that 
 another 3-face $[v_3'v_4'v_5']$ exists;
$\{[v_1'v_2'v_3'],[v_3'v_4'v_5']\}$ is a chain.
Moreover, 
 observe that
if $v_1' \neq v_c$ also holds, then $[v_{t-1}' v_t' v_1']$ is a 3-face
and
$\{[v_{t-1}' v_t' v_1'],[v_1'v_2'v_3'],[v_3'v_4'v_5']\}$ is a chain.
By repeating similar observations, it follows  that
there exists a chain ${\cal C}=\{[v_1'v_2'v_3'],[v_3'v_4'v_5'],\ldots,
[v_{t-1}' v_{t}' v_1']\}$.
Then notice that $d_{G'}(v_i')=4$ (resp., 2) if $i$ is odd (resp., even);
by $\Delta(G)=4$, 
  if $v_c$ exists, then
 $v_c$ is some vertex $v_j'$ with $d_{G'}(v_j')=2$. 
Hence, we can see that if 
$v_c$ exists, then 
 $G'-v_c$ contains a  chain with an inner edge $(v'_{j-1},v_{j+1}')$,
and that otherwise ${\cal C}-\{[v_{t-1}' v_t' v_1']\}$
is a  chain with an inner edge $(v_1',v_{t-1}')$ as required.

(Case-II) For an inner edge $(v_i',v_j') \in E$, 
let $G[v_i',v_j']$ denote the subgraph of $G$ induced by
$\{v_i',v_{i+1}',\ldots,v_{j-1}', $ $v_j'\}$.
Without loss of generality, 
let $(v'_1,v'_i)$ be an inner edge in $G'$ which 
does not belong to any endface such that $G[v'_2,v'_{i-1}]$
does not contain $v_c$.
Moreover, we can choose such an edge $(v'_1,v'_i)$  that any inner edge
in $G[v'_1,v'_i]$ belongs to some endface of $G$.
Note that 
if $G[v_1',v_i']$ contains an inner edge $e'=(v_j',v_k')$ $(1 \leq
 j\leq k\leq i)$ 
not in any endface, then
the number of  inner edges not in any endface in $G[v_j',v_k']$
is less than that in $G[v_1',v_i']$ and this observation
indicates that this choice of  $(v_1',v_i')$  is possible.
Then by applying the similar arguments in Case-I to  endfaces
in $G[v'_1,v'_i]$ (also in $G$), we can observe that
there exists a chain $\{[v_1'v_2'v_3'],[v_3'v_4'v_5'],\ldots,
[v_{i-2}'v_{i-1}'v_i']\}$, which is a required chain. 
\qed
\end{proof}

\nop{
In \cite{CW07}, Chen and Wang showed that 
$\lambda^T_2(G) \leq \Delta(G)+2$ if $G$ does not contain the
configuration (C3).
We here consider the case where $G$ contains the configuration
}

~

By utilizing this lemma, we show that 
$\lambda_2^T(G)\leq \Delta(G)+2$ holds even if $\Delta(G)=4$.

\begin{theorem}\label{delta4:theo}
If $G=(V,E)$ is an outerplane graph with $\Delta(G)=4$, 
then $\lambda^T_2(G)\leq 6$. 
\end{theorem}
\begin{proof}
We prove this by induction on $k=|V(G)|+|E(G)|$.
The theorem clearly holds if $k \leq 9$.
Consider the case of $k \geq 10$ and assume that
for each $k' < k$, this theorem holds. 
We also assume that $G$ is connected, since otherwise
we can treat  each component separately.
Thus, $1 \leq \delta(G) \leq 2$.

In the case of $\delta(G)=1$, similarly to 
the proof of Theorem~\ref{delta3:theo},
we can observe that $G$ has 
 a (2,1)-total labeling $f: V(G)
 \cup E(G) \to {\cal L}_6$.

\nop{
Consider the case where $\delta(G)=1$.
Let $u_1$ be a vertex with $d(u_1)=1$.
By the induction hypothesis, $G-u_1$ has a (2,1)-total labeling $f: V(G-u_1)
 \cup E(G-u_1) \to {\cal L}_6$.
Let $u_2$ be the neighbor of $u_1$ in $G$ and
$u_3,u_4,u_5$ be vertices adjacent to $u_2$ in $G-u_1$
where $u_i=u_j$ may occur for $\{i,j\}\subseteq \{3,4,5\}$ 
 (note that $\Delta(G)=4$).
Hence we can extend $f$ to the edge $(u_1,u_2)$ and the vertex $u_1$ 
as follows:
 assign a label $a \in$
  ${\cal L}_6-$ $\{f(u_2)-1,f(u_2),f(u_2)+1,f(u_2 u_3), f(u_2 u_4)$,
 $f(u_2 u_5)\}$
to  $(u_1,v_1)$, and then
a label in ${\cal L}_6-\{f(u_2),a-1,a,a+1\}$ to $u_1$.
}

Consider the case where $\delta(G)=2$.
By Theorem~\ref{config:theo}, $G$ contains at least one of 
the configurations (C1)--(C3).
In \cite[Theorem 13]{CW07}, Chen and Wang showed that
if $G$ contains (C1) or (C2), then  
a (2,1)-total labeling of $G$ can be obtained from a
feaible labeling of 
some subgraph $H$ of $G$.
Here we  only consider the case where $G$ contains neither (C1) nor (C2).
By Lemma~\ref{config2:lem}, $G$ contains
a chain  ${\cal C}=\{[u_1 u_2 u_3],[u_3 u_4 u_5],\ldots,[u_{2t-1} u_{2t} u_{2t+1}]\}$
 of 3-faces such that $(u_1,u_{2t+1})$ is an inner edge.
 Let $w_1$ (resp., $w_{2t+1}$) denote the vertex not in ${\cal C}$ 
which is adjacent
to $u_1$ (resp., $u_{2t+1}$) (note that
by $\Delta(G)=4$, no edge other than
$(u_1,w_1)$ and $(u_{2t+1},w_{2t+1})$
connects 
${\cal C}$ and the remaining part of $G$).
By the induction hypothesis,
 $H:=(G-\{u_2,u_3,\ldots,u_{2t}\})-\{(u_1,u_{2t+1})\}$
has a (2,1)-total labeling $f: V(H)
 \cup E(H) \to {\cal L}_6$. 
Then by $d_{H}(u_1)=d_{H}(u_{2t+1})=1$, 
we can assign    any label 
 in ${\cal L}_6-\{f(w_1),f(u_1 w_1)-1,f(u_1 w_1),f(u_1 w_1)+1\}$
(resp., ${\cal L}_6-\{f(w_{2t+1}),f(u_{2t+1} w_{2t+1})-1,f(u_{2t+1}
 w_{2t+1}),
f(u_{2t+1} w_{2t+1})+1\}$)
to $u_1$ (resp., $u_{2t+1}$)  without violating the feasibility.
Let $L(u_1;f)$ and $L(u_{2t+1};f)$ denote
the sets of such possible labels with respect to $f$
for $u_1$ and $u_{2t+1}$, respectively;  
$L(u_1;f)={\cal L}_6-\{f(w_1), $ $f(u_1 w_1)$ $-1,f(u_1 w_1),f(u_1 w_1)+1\}$
and
$L(u_{2t+1};f)={\cal L}_6-\{f(w_{2t+1}),f(u_{2t+1} w_{2t+1})-1, $ $f(u_{2t+1}
 w_{2t+1}),$ $
f(u_{2t+1} w_{2t+1})+1\}$.
Notice that $|L(u_1;f)|\geq 3$ and $|L(u_{2t+1};f))|\geq 3$.

\begin{claim}
$L(u_1;f)\times L(u_{2t+1};f)$ contains one of
$(0,6),(0,1),$ $(1,2),(2,3),(3,$
 $4),(4,5),(5,$ $6),$ $(6,0),(6,5),(5,4),(4,$ $3),$
$(3,2),(2,1)$, and $(1,0)$.
\end{claim}
\begin{proof}
Assume for contradiction that this claim does not hold.
First consider the case of $\{0,6\} \subseteq L(u_1;f)$.
By assumption, $\{0,6\} \cap  L(u_{2t+1};f)=\emptyset$.
This inidcates  that (a1) $f(w_{2t+1})=0$ and $f(u_{2t+1} w_{2t+1})\in \{5,6\}$
or (a2)
 $f(w_{2t+1})=6$ and $f(u_{2t+1} w_{2t+1})\in \{0,1\}$.
In the case of (a1), we have
 $1 \in L(u_{2t+1};f)$ (i.e., $(0,1) \in L(u_1;f)\times
 L(u_{2t+1};f))$ and in the case of (a2), we have
$5 \in L(u_{2t+1};f)$ (i.e., $(6,5) \in L(u_1;f)\times
 L(u_{2t+1};f))$), a contradiction.

Next consider the case of $0 \in L(u_1;f)$ and $6 \notin L(u_1;f)$.
By assumption, $\{1,6\}\cap L(u_{2t+1};f)=\emptyset$.
This indicates that (b1)
 $f(w_{2t+1})=1$ and $f(u_{2t+1} w_{2t+1})=5$,
(b2)
 $f(w_{2t+1})=1$ and $f(u_{2t+1} w_{2t+1})=6$,
(b3)
 $f(w_{2t+1})=6$ and $f(u_{2t+1} w_{2t+1})=0$,
(b4)
  $f(w_{2t+1})=6$ and $f(u_{2t+1} w_{2t+1})=1$,
or
(b5)
  $f(w_{2t+1})=6$ and $f(u_{2t+1} w_{2t+1})=2$.
In the case of (b1) (resp., (b2), (b3), (b4), (b5)), 
we have $L(u_{2t+1};f)=\{0,2,3\}$
(resp., $\{0,2,3,4\}$, $\{2,3,4,5\}$, $\{3,4,5\}$, $\{0,4,5\}$).
By assumption that $f$ 
does not satisfy this claim, it follows that 
in the case of (b1) (resp., (b2), (b3), (b4), (b5)), 
$L(u_1;f) \subseteq \{0,5\}$
(resp., $\{0\}$, $\{0\}$ $\{0,1\}$, $\{0,2\}$).
All of these cases contraict $|L(u_1;f)|\geq 3$.

By symmetry of labelings, the case of $6 \in L(u_1;f)$ and $0 \notin L(u_1;f)$
can be treated similarly to the previous case.
Finally, consider the case of $\{0,6\} \cap L(u_1;f)=\emptyset$.
Assume that  $\{0,6\} \cap L(u_{2t+1};f)=\emptyset$,
since any other case  can be treated
similarly by exchanging parts of $u_1$ and $u_{2t+1}$.
Then we can observe that 
$3 \in L(u_1;f)$, $\{2,4\}\cap L(u_1;f)\neq \emptyset$,
$3 \in L(u_{2t+1};f)$, and $\{2,4\}\cap L(u_{2t+1};f)\neq \emptyset$.
It follows that 
$(2,3)$ or $(4,3)$ are contained in  $L(u_1;f)\times L(u_{2t+1};f)$,
a contradiction. \qed
\end{proof}

~

By symmetry, it suffices to consider the following four cases:
(Case-1) $0 \in L(u_1;f)$ and $6 \in L(u_{2t+1};f)$,
(Case-2) $0 \in L(u_1;f)$ and $1 \in L(u_{2t+1};f)$,
(Case-3) $1 \in L(u_1;f)$ and $2 \in L(u_{2t+1};f)$, and
(Case-4) $2 \in L(u_1;f)$ and $3 \in L(u_{2t+1};f)$.
In each case, we will extend $f$ to  
a (2,1)-total labeling $f':V\cup E \to {\cal L}_6$ of $G$.
Let $f'(x):=f(x)$ for every  $x \in V(H) \cup E(H)$.

(Case-1) Let $f'(u_1):=0$ and $f'(u_{2t+1}):=6$.

(1.1) Assume that $t$ is even; $t=2k$. 

(1.1.1) Assume that $f'(u_1 w_1)\neq 6$ and $f'(u_{2t+1} w_{2t+1})\neq
 0$.

Let 
$f'(u_2):=6$,
$f'(u_3):=3$,
$f'(u_4):=0$,
$f'(u_5):=6$,
$f'(u_2 u_3):=1$,
$f'(u_1u_3):=6$,
$f'(u_3u_4):=5$,
$f'(u_4u_5):=4$, and
$f'(u_3u_5):=0$.
For $i=2,3,\ldots,k$, let
$f'(u_{4i-2}):=1$,
$f'(u_{4i-1}):=3$,
$f'(u_{4i}):=0$,
$f'(u_{4i+1}):=6$,
$f'(u_{4i-3}u_{4i-2}):=3$,
 $f'(u_{4i-2}u_{4i-1}):=6$,
$f'(u_{4i-3}u_{4i-1}):=1$,
$f'(u_{4i-1}u_{4i}):=5$,
$f'(u_{4i}u_{4i+1}):=4$, and
$f'(u_{4i-1}u_{4i+1})
$ $:=0$.
Denote this labeling $f'$ by $f_1$ (see Fig.~\ref{fig:f1}).
\begin{figure*}[ht]
  \begin{center}
   \resizebox{3.6truein}{2.88truein}{\includegraphics{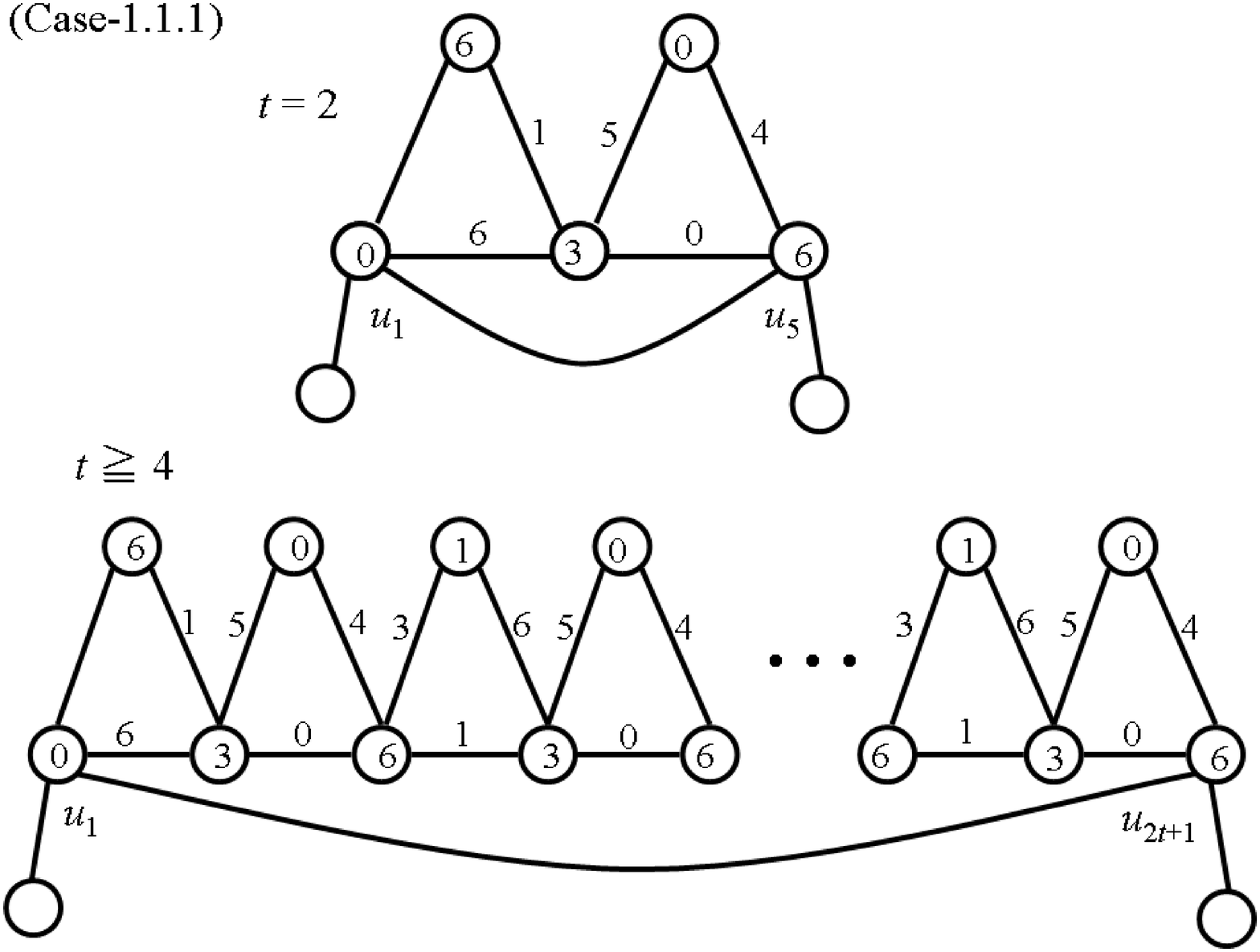}}
  \end{center}
 \caption{A labeling $f_1$ on ${\cal C}$.}
 \label{fig:f1}
\end{figure*}
We next assign a label $\{2,3,4\}-\{f'(u_1w_1),f'(u_{2t+1}w_{2t+1})\}$
to $(u_1,u_{2t+1})$, and
a label in
$\{2,3,4\}-\{f'(u_1w_1),f'(u_1u_{2t+1})\}$ to $(u_1,u_2)$.
Reassign 
a label in 
$\{2,3,4\}-\{f'(u_{2t+1}w_{2t+1}), $ $f'(u_1u_{2t+1})\}$ to
 $(u_{2t},u_{2t+1})$.
Then the resulting labeling $f'$ is a 
a (2,1)-total labeling of $G$.

(1.1.2) Assume that $f'(u_1 w_1)= 6$ and $f'(u_{2t+1} w_{2t+1})\neq
 0$.

Let $f':=f_1$.
We reassign labels for some vertices and edges as follows.
Let $f'(u_2):=1$, $f'(u_3):=2$,
 $f'(u_1u_2):=5$, $f'(u_2 u_3):=6$
and  $f'(u_1 u_3):=4$.
Next we assign a label in $\{2,3\}-\{f'(u_{2t+1}w_{2t+1})\}$ to
$(u_1, u_{2t+1})$
and reassign a label
in $\{2,3,4\}-\{f'(u_1u_{2t+1}),f'(u_{2t+1}w_{2t+1})\}$
to 
 $(u_{2t},u_{2t+1})$.
Then the resulting labeling $f'$ is a 
a (2,1)-total labeling of $G$.


(1.1.3) Assume that $f'(u_1 w_1)\neq 6$ and $f'(u_{2t+1} w_{2t+1})=
 0$.

Let $f':=f_1$.
We reassign labels for some vertices and edges as follows.
Let 
$f'(u_{2t-1}):=4$,  $f'(u_{2t}):=5$,
$f'(u_{2t-1}u_{2t}):=0$,
$f'(u_{2t}u_{2t+1}):=1$, and $f'(u_{2t-1}u_{2t+1}):=2$.
Next we assign a label in $\{3,4\}-\{f'(u_1w_1)\}$ to
$(u_1, u_{2t+1})$
and 
 a label in
$\{2,3,4\}-\{f'(u_1u_{2t+1}),f'(u_1w_1)\}$ to $(u_1,u_2)$.
Then the resulting labeling $f'$ is a 
 (2,1)-total labeling of $G$.


(1.1.4) Assume that $f'(u_1 w_1)= 6$ and $f'(u_{2t+1} w_{2t+1})=
 0$.

We can obtain a (2,1)-total labeling $f'$ of $G$ as follows.
If $t=2$, then 
let 
$f'(u_2):=3$,
$f'(u_3):=1$,
$f'(u_4):=3$,
$f'(u_1u_2):=5$,
$f'(u_2u_3):=6$,
$f'(u_1u_3):=4$,
$f'(u_3u_4):=5$,
 $f'(u_4u_5):=1$, 
$f'(u_3u_5)=3$,
and
$f'(u_1u_5):=2$.

Assume that $t\geq 4$.
Let $f':=f_1$.
We reassign labels for some vertices and edges similarly to
(1.1.2) and (1.1.3).
Namely, 
let $f'(u_2):=1$, $f'(u_3):=2$,
$f'(u_1u_2):=5$, $f'(u_2u_3):=6$, $f'(u_1u_3):=4$,
$f'(u_{2t-1}):=4$,  $f'(u_{2t}):=5$,
 $f'(u_{2t-1}u_{2t}):=0$,
$f'(u_{2t}u_{2t+1}):=1$, $f'(u_{2t-1}u_{2t+1}):=2$,
and $f'(u_1u_{2t+1}):=3$.

(1.2) Assume that $t$ is odd; $t=2k+1~(k\geq 1)$ . 

(1.2.1) Assume that $f'(u_1 w_1)\neq 6$ and $f'(u_{2t+1} w_{2t+1})\neq
 0$.

Let $f'(u_2):=6$,
$f'(u_3):=3$,
$f'(u_4):=0$,
$f'(u_5):=4$,
$f'(u_6):=0$,
 $f'(u_7):=6$, 
$f'(u_2 u_3):=0$,
$f'(u_1u_3):=6$, 
$f'(u_3u_4):=5$,
 $f'(u_4u_5):=2$,
$f'(u_3u_5):=1$,
$f'(u_5u_6):=6$, 
$f'(u_6u_7):=4$, and
$f'(u_5 u_7):=0$.
For $i=2,3,\ldots,k$,
let 
$f'(u_{4i}):=0$,
  $f'(u_{4i+1}):=4$, 
$f'(u_{4i+2}):=0$, 
$f'(u_{4i+3}):=6$,
 $f'(u_{4i-1}u_{4i}):=3$,
 $f'(u_{4i}u_{4i+1}):=2$, 
$f'(u_{4i-1}u_{4i+1}):=1$,
 $f'(u_{4i+1}u_{4i+2}):=6$,
$f'(u_{4i+2}u_{4i+3}):=4$, and
$f'(u_{4i+1}u_{4i+3}):=0$.
Denote this labeling $f'$ by $f_2$ (see Fig.~\ref{fig:f2}).
\begin{figure*}[ht]
  \begin{center}
   \resizebox{3.6truein}{2.88truein}{\includegraphics{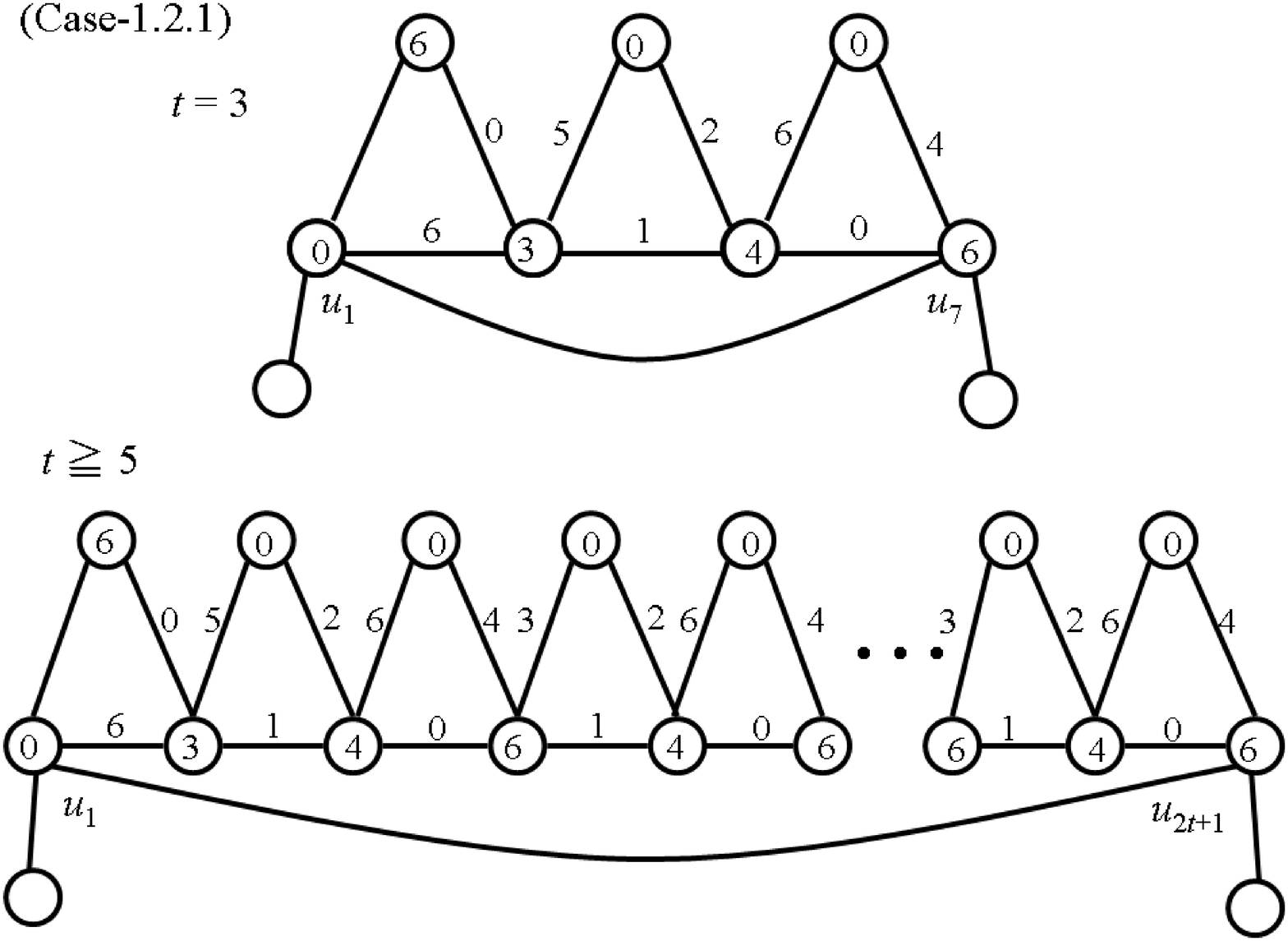}}
  \end{center}
 \caption{A labeling $f_2$ on ${\cal C}$.}
 \label{fig:f2}
\end{figure*}
We next assign a label $\{2,3,4\}-\{f'(u_1w_1),f'(u_{2t+1}w_{2t+1})\}$
to $(u_1,u_{2t+1})$ and
assign 
a label in
$\{2,3,4\}-\{f'(u_1w_1),f'(u_1u_{2t+1})\}$ to $(u_1,u_2)$.
Reassign
a label in 
$\{2,3,4\}-\{f'(u_{2t+1}w_{2t+1}),f'(u_1u_{2t+1})\}$ to
 $(u_{2t},u_{2t+1})$.
%

(1.2.2) Assume that $f'(u_1 w_1)= 6$ and $f'(u_{2t+1} w_{2t+1})\neq
 0$.

Let $f':=f_2$.
We reassign labels for some vertices and edges as follows.
Let $f'(u_2):=2$, $f'(u_3):=6$,
 $f'(u_1u_2):=5$,
$f'(u_1 u_3):=4$,
and 
$f'(u_3u_4):=3$.
Next we assign a label in $\{2,3\}-\{f'(u_{2t+1}w_{2t+1})\}$ to
$(u_1, u_{2t+1})$.
Reassign a label in
$\{2,3,4\}-\{f'(u_1u_{2t+1}),f'(u_{2t+1}w_{2t+1})\}$ to $(u_{2t},u_{2t+1})$.

(1.2.3) Assume that $f'(u_1 w_1)\neq 6$ and $f'(u_{2t+1} w_{2t+1})=
 0$.

Let $f':=f_2$.
We reassign labels for some vertices and edges as follows.
Let  $f'(u_{2t-1}):=5$,
$f'(u_{2t}):=3$,
$f'(u_{2t-1}u_{2t}):=0$,
$f'(u_{2t}u_{2t+1}):=1$, and 
$f'(u_{2t-1}u_{2t+1}):=3$.
Next we assign a label in $\{2,4\}-\{f'(u_1w_1)\}$ to
$(u_1, u_{2t+1})$ and
 a label in
$\{2,3,4\}-\{f'(u_1w_1),f'(u_1u_{2t+1})\}$ to $(u_1,u_2)$.

(1.2.4) Assume that $f'(u_1 w_1)= 6$ and $f'(u_{2t+1} w_{2t+1})=
 0$.

Let $f':=f_2$.
We reassign labels for some vertices and edges similarly to
(1.2.2) and (1.2.3).
Namely, 
let $f'(u_2):=2$, $f'(u_3):=6$,
$f'(u_1u_2):=5$,
$f'(u_1u_3):=4$,
$f'(u_3u_4):=3$,
 $f'(u_{2t-1}):=5$,
$f'(u_{2t}):=3$,
$f'(u_{2t-1}u_{2t}):=0$,
$f'(u_{2t}u_{2t+1}):=1$, 
$f'(u_{2t-1}u_{2t+1}):=3$,
 and 
$f'(u_1u_{2t+1}):=2$.

(Case-2) Let $f'(u_1):=0$ and $f'(u_{2t+1}):=1$.

(2.1) Assume that $t$ is even; $t=2k$. 

(2.1.1) Assume that $f'(u_1 w_1)\neq 2$ and $f'(u_{2t+1} w_{2t+1})\notin
\{3,4\}$.

Let 
$f'(u_2):=2$,
$f'(u_3):=6$,
$f'(u_4):=5$,
$f'(u_5):=1$,
$f'(u_2 u_3):=0$,
$f'(u_1u_3):=2$,
$f'(u_3u_4):=1$,
$f'(u_4u_5):=3$, and
$f'(u_3u_5):=4$.
For $i=2,3,\ldots,k$, let
$f'(u_{4i-2}):=4$,
$f'(u_{4i-1}):=0$,
$f'(u_{4i}):=2$,
$f'(u_{4i+1}):=1$,
$f'(u_{4i-3}u_{4i-2}):=6$,
 $f'(u_{4i-2}u_{4i-1}):=2$,
$f'(u_{4i-3}u_{4i-1}):=5$,
$f'(u_{4i-1}u_{4i}):=6$,
$f'(u_{4i}u_{4i+1}):=4$, and
$f'(u_{4i-1}u_{4i+1})$
$:=3$.
Denote this labeling $f'$ by $f_3$ (see Fig.~\ref{fig:f3}).
\begin{figure*}[ht]
  \begin{center}
   \resizebox{3.6truein}{2.88truein}{\includegraphics{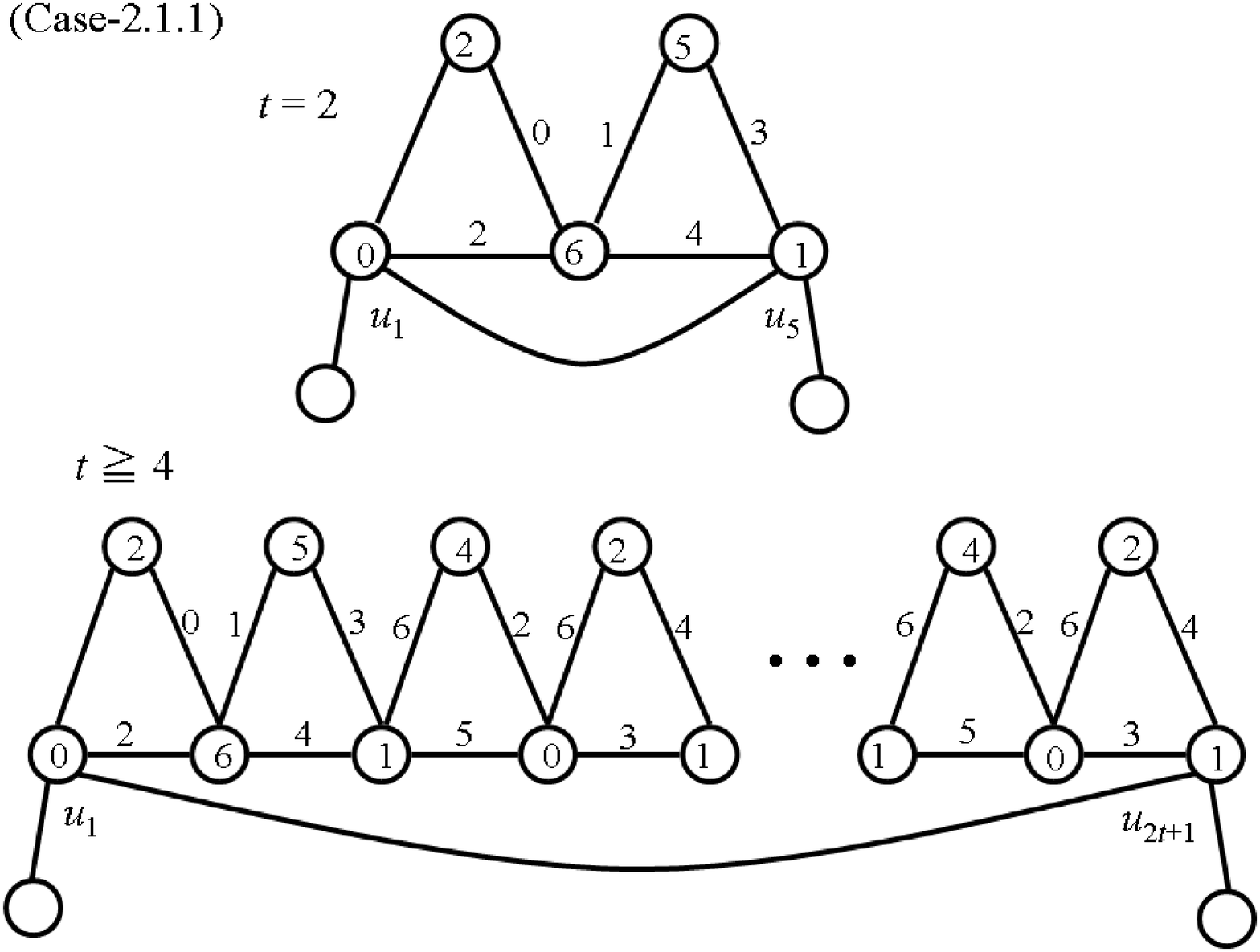}}
  \end{center}
 \caption{A labeling $f_3$ on ${\cal C}$.}
 \label{fig:f3}
\end{figure*}

Consider the case of $t=2$.
We  assign a label $\{3,5,6\}-\{f'(u_1w_1),f'(u_5w_5)\}$
to $(u_1,u_5)$ and
 a label in 
$\{4,5,6\}-\{f'(u_1w_1),f'(u_1u_5)\}$ to $(u_1,u_2)$.
Reassign a label in 
$\{3,5,6\}-\{f'(u_5w_5),f'(u_1u_5)\}$ to $(u_4,u_5)$.
Then if $f'(u_4u_5)\in \{5,6\}$, then reassign
a label for $u_4$ as
$f'(u_4):=3$.

Consider the case of  $t \geq 4$.
Assign a label $\{4,5,6\}-\{f'(u_1w_1),f'(u_{2t+1}w_{2t+1})\}$
to $(u_1,u_{2t+1})$ and
 a label in 
$\{4,5,6\}-\{f'(u_1w_1),f'(u_1u_{2t+1})\}$ to $(u_1,u_2)$.
Reassign
a label in 
$\{4,5,6\}-\{f'(u_{2t+1}w_{2t+1}),f'(u_1u_{2t+1})\}$ to $(u_{2t},u_{2t+1})$.
Then if $f'(u_{2t} u_{2t+1})=6$, then reassign
a label for $(u_{2t-1},u_{2t})$ as $f'(u_{2t-1}u_{2t}):=4$.

(2.1.2) Assume that $f'(u_1 w_1)= 2$ and $f'(u_{2t+1} w_{2t+1})\notin
 \{3,4\}$.

Let $f':=f_3$.
We reassign a label for $(u_1,u_3)$ to 3.

Consider the case of $t=2$.
Assign a label in $\{5,6\}-\{f'(u_5w_5)\}$ to
$(u_1, u_5)$ and
a label in $\{4,5,6\}-\{f'(u_1u_5)\}$
to $(u_1,u_2)$.
Reassign 
a label in
$\{3,5,6\}-\{f'(u_5w_5),f'(u_1u_5)\}$ to $(u_4,u_5)$.
Then if $f'(u_4u_5)\in \{5,6\}$, then let
$f'(u_4):=3$.


Consider the case of $t\geq 4$.
Assign a label in $\{4,5,6\}-\{f'(u_{2t+1}w_{2t+1})\}$ to
$(u_1, u_{2t+1})$ and
a label in $\{4,5,6\}$ $-\{f'(u_1u_{2t+1})\}$
to $(u_1,u_2)$.
Reassign 
a label in
$\{4,5,6\}-\{f'(u_{2t+1}w_{2t+1}),$ 
$f'(u_1u_{2t+1})\}$ to $(u_{2t},u_{2t+1})$.
Then, if $f'(u_{2t}u_{2t+1})$ $=6$, then let
$f'(u_{2t-1}u_{2t}):=4$.



(2.1.3) Assume that $f'(u_1 w_1)\neq 2$ and $f'(u_{2t+1} w_{2t+1})=3$.

If $t=2$, then we can obtain a feasible labeling similarly to
(2.1.1).
Consider the case of $t\geq 4$.
Let $f':=f_3$.
Reassign a label for $(u_{2t-1},u_{2t+1})$ as $f'(u_{2t-1}u_{2t+1}):=6$.
Next we assign a label in $\{4,5\}-\{f'(u_1w_1)\}$ to
$(u_1, u_{2t+1})$
and  
 a label in
$\{4,5,6\}-\{f'(u_1u_{2t+1}),f'(u_1w_1)\}$ to $(u_1,u_2)$.
Reassign a label in 
$\{4,5\}-\{f'(u_1u_{2t+1})\}$ to $(u_{2t},u_{2t+1})$.
Then if $f'(u_{2t}u_{2t+1})=4$, then
let $f'(u_{2t}):=6$ and $f'(u_{2t-1}u_{2t}):=3$, and
if $f'(u_{2t}u_{2t+1})=5$, then
let  $f'(u_{2t-1}u_{2t}):=4$.


(2.1.4) Assume that $f'(u_1 w_1)\neq 2$ and $f'(u_{2t+1} w_{2t+1})=4$.

If $t\geq 4$, then we can obtain a feasible labeling similarly to
(2.1.1).
Consider the case of $t=2$.
Let $f':=f_3$.
We reassign labels for $(u_3,u_5)$ and $u_4$ 
as $f'(u_3u_5):=3$ and $f'(u_4):=3$.
Next we assign a label in $\{5,6\}-\{f'(u_1w_1)\}$ to
$(u_1, u_5)$
and 
 a label in
$\{4,5,6\}-\{f'(u_1u_5),f'(u_1w_1)\}$ to $(u_1,u_2)$.
Reassign a label in 
$\{5,6\}-\{f'(u_1u_5)\}$ to $(u_4,u_5)$.

(2.1.5) Assume that $f'(u_1 w_1)= 2$ and $f'(u_{2t+1} w_{2t+1})=3$.

If $t=2$, then we can obtain a feasible labeling similarly to
(2.1.2).
Consider the case of $t\geq 4$.
Let $f':=f_3$.
We reassign labels for some vertices and edges similarly to
(2.1.2) and (2.1.3).
Namely, 
let $f'(u_1u_2):=4$,
$f'(u_1u_3):=3$,
$f'(u_{2t}):=6$,
$f'(u_{2t-1}u_{2t}):=3$,
$f'(u_{2t}u_{2t+1}):=4$, $f'(u_{2t-1}u_{2t+1}):=6$, and
$f'(u_1u_{2t+1}):=5$.

(2.1.6) Assume that $f'(u_1 w_1)= 2$ and $f'(u_{2t+1} w_{2t+1})=4$.

If $t\geq 4$, then we can obtain a feasible labeling similarly to
(2.1.2).
If $t=2$,
let 
$f'(u_2):=2$,
$f'(u_3):=6$,
$f'(u_4):=3$,
$f'(u_1u_2):=5$,
$f'(u_2u_3):=0$,
$f'(u_1u_3):=4$,
$f'(u_3u_4):=1$,
 $f'(u_4u_5):=5$, 
$f'(u_3u_5)=3$,
and
$f'(u_1u_5):=6$.


(2.2) Assume that $t$ is odd; $t=2k+1~(k\geq 1)$ . 

(2.2.1) Assume that $f'(u_1 w_1)\neq 2$ and $f'(u_{2t+1} w_{2t+1})\neq
 5$.

Let $f'(u_2):=2$,
$f'(u_3):=6$,
$f'(u_4):=4$,
$f'(u_5):=0$,
$f'(u_6):=6$,
 $f'(u_7):=1$, 
$f'(u_2 u_3):=0$,
$f'(u_1u_3):=2$, 
$f'(u_3u_4):=1$,
 $f'(u_4u_5):=6$,
$f'(u_3u_5):=4$,
$f'(u_5u_6):=2$, 
$f'(u_6u_7):=3$, and
$f'(u_5 u_7):=5$.
For $i=2,3,\ldots,k$,
let 
$f'(u_{4i}):=6$,
  $f'(u_{4i+1}):=0$, 
$f'(u_{4i+2}):=6$, 
$f'(u_{4i+3}):=1$,
 $f'(u_{4i-1}u_{4i}):=4$,
 $f'(u_{4i}u_{4i+1}):=3$, 
$f'(u_{4i-1}u_{4i+1}):=6$,
 $f'(u_{4i+1}u_{4i+2}):=2$,
$f'(u_{4i+2}u_{4i+3}):=3$, and
$f'(u_{4i+1}u_{4i+3}):=5$.
Denote this labeling $f'$ by $f_4$ (see Fig.~\ref{fig:f4}).
\begin{figure*}[ht]
  \begin{center}
   \resizebox{3.6truein}{2.88truein}{\includegraphics{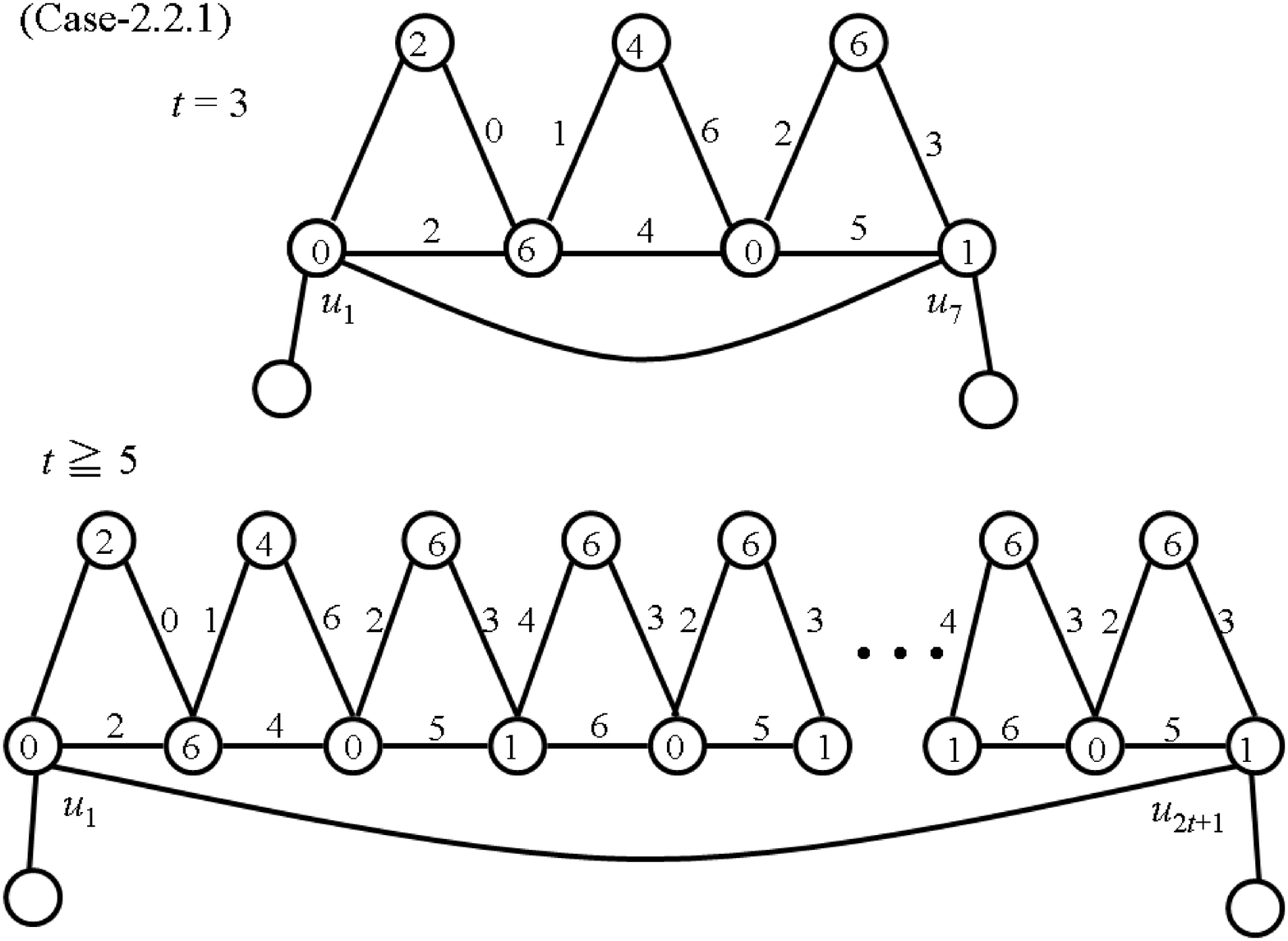}}
  \end{center}
 \caption{A labeling $f_4$ on ${\cal C}$.}
 \label{fig:f4}
\end{figure*}
We next assign a label $\{3,4,6\}-\{f'(u_1w_1),f'(u_{2t+1}w_{2t+1})\}$
to $(u_1,u_{2t+1})$.
and
a label in
$\{4,5,6\}-\{f'(u_1w_1),f'(u_1u_{2t+1})\}$ to $(u_1,u_2)$.
Reassign
a label in 
$\{3,4,6\}-\{f'(u_{2t+1}w_{2t+1}),f'(u_1u_{2t+1})\}$ to
 $(u_{2t},u_{2t+1})$.
Then if $f'(u_{2t}u_{2t+1})=6$,
then let $f'(u_{2t}):=4$.
%

(2.2.2) Assume that $f'(u_1 w_1)= 2$ and $f'(u_{2t+1} w_{2t+1})\neq
 5$.

Let $f':=f_4$.
We reassign labels for some vertices and edges as follows.
Let $f'(u_1u_2):=5$ and
$f'(u_1 u_3):=3$.
Next we assign a label in $\{4,6\}-\{f'(u_{2t+1}w_{2t+1})\}$ to
$(u_1, u_{2t+1})$.
Reassign a label in
$\{3,4,6\}-\{f'(u_1u_{2t+1}), $ $f'(u_{2t+1}w_{2t+1})\}$ to $(u_{2t},u_{2t+1})$.
Then if $f'(u_{2t}u_{2t+1})=6$,
then let $f'(u_{2t}):=4$.

(2.2.3) Assume that $f'(u_1 w_1)\neq 2$ and $f'(u_{2t+1} w_{2t+1})=
 5$.

Let $f':=f_4$.

First consider the case of $t=3$.
Let $f'(u_6):=2$,
$f'(u_5u_6):=5$, and
$f'(u_5u_7):=3$.
Next we assign a label in $\{4,6\}-\{f'(u_1w_1)\}$ to
$(u_1, u_7)$ and
 a label in
$\{4,5,6\}-\{f'(u_1w_1),f'(u_1u_7)\}$ to $(u_1,u_2)$.
Reassign
a label in
$\{4,6\}-\{f'(u_1u_7)\}$ to $(u_6,u_7)$.

Next consider the case of $t\geq 5$.
Let 
$f'(u_{2t-1}u_{2t+1}):=4$.
Next we assign a label in $\{3,6\}-\{f'(u_1w_1)\}$ to
$(u_1, u_{2t+1})$ and 
 a label in
$\{4,5,6\}-\{f'(u_1w_1),f'(u_1u_{2t+1})\}$ to $(u_1,u_2)$.
Reassign
a label in
$\{3,6\}-\{f'(u_1u_{2t+1})\}$ to $(u_{2t},u_{2t+1})$.
Then if $f'(u_{2t}u_{2t+1})=6$, then
$f'(u_{2t}):=4$.

(2.2.4) Assume that $f'(u_1 w_1)= 2$ and $f'(u_{2t+1} w_{2t+1})=
 5$.

Let $f':=f_4$.
We reassign labels for some vertices and edges similarly to
(2.2.2) and (2.2.3).
Namely, if $t=3$,
 let $f'(u_1u_2):=5$,
$f'(u_1u_3):=3$,
$f'(u_6):=2$,
$f'(u_5u_6):=5$,
$f'(u_6u_7):=6$,
$f'(u_5u_7):=3$, and
$f'(u_1u_7):=4$.
If $t\geq 5$, then
let $f'(u_1u_2):=5$,
$f'(u_1u_3):=3$,
 $f'(u_{2t-1}u_{2t+1}):=4$, and
$f'(u_1u_{2t+1}):=6$.

(Case-3) Let $f'(u_1):=1$ and $f'(u_{2t+1}):=2$.

(3.1) Assume that $t$ is even; $t=2k$. 

(3.1.1) Assume that $f'(u_1 w_1)\neq 3$ and $f'(u_{2t+1} w_{2t+1})\neq 0$.

Let 
$f'(u_2):=0$,
$f'(u_3):=5$,
$f'(u_4):=3$,
$f'(u_5):=2$,
$f'(u_2 u_3):=2$,
$f'(u_1u_3):=3$,
$f'(u_3u_4):=1$,
$f'(u_4u_5):=6$, and
$f'(u_3u_5):=0$.
For $i=2,3,\ldots,k$, let
$f'(u_{4i-2}):=0$,
$f'(u_{4i-1}):=6$,
$f'(u_{4i}):=0$,
$f'(u_{4i+1}):=2$,
$f'(u_{4i-3}u_{4i-2}):=5$,
 $f'(u_{4i-2}u_{4i-1}):=3$,
$f'(u_{4i-3}u_{4i-1}):=4$,
$f'(u_{4i-1}u_{4i}):=2$,
$f'(u_{4i}u_{4i+1}):=6$, and
$f'(u_{4i-1}u_{4i+1})$ $:=0$.
Denote this labeling $f'$ by $f_5$ (see Fig.~\ref{fig:f5}).
\begin{figure*}[ht]
  \begin{center}
   \resizebox{3.6truein}{2.88truein}{\includegraphics{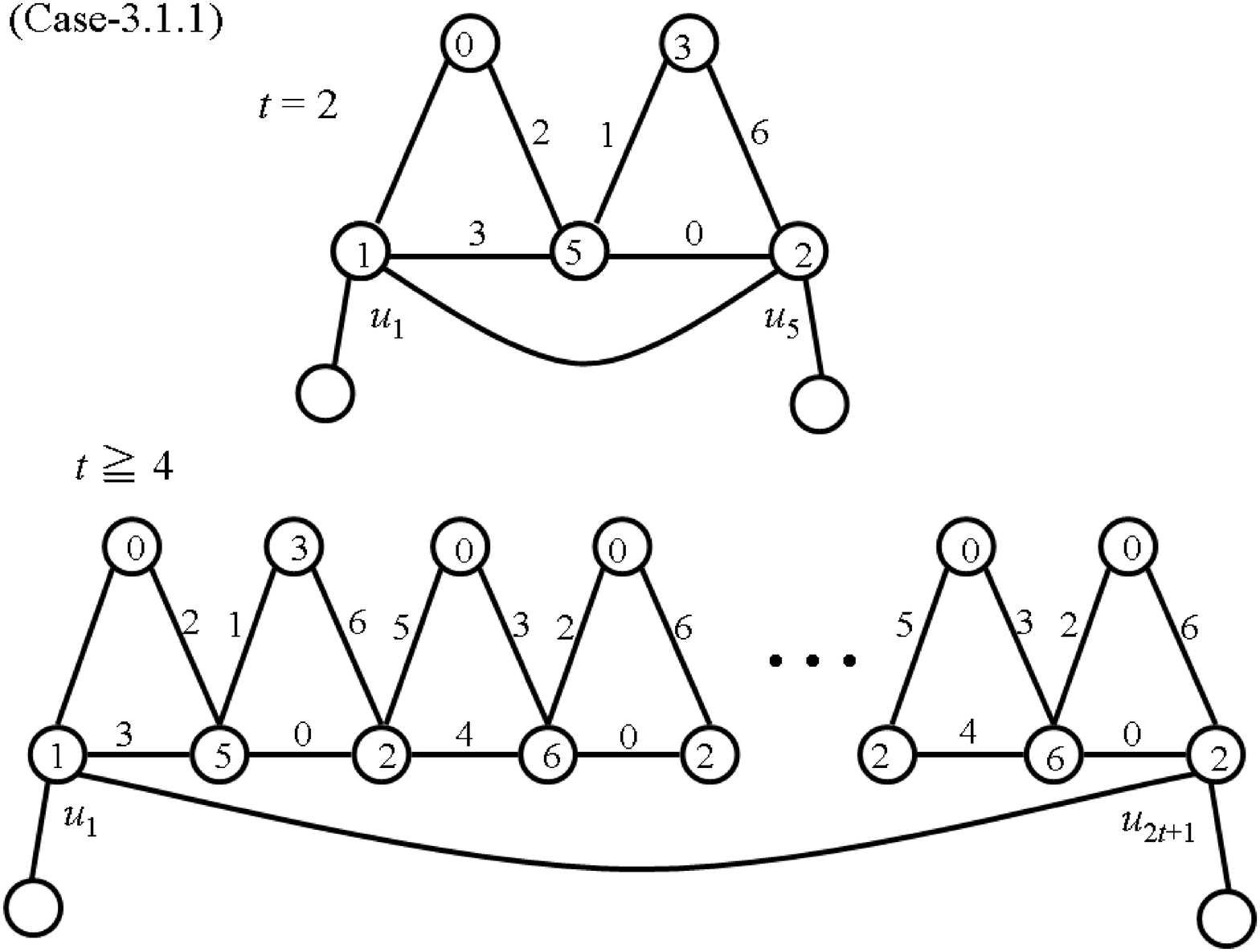}}
  \end{center}
 \caption{A labeling $f_5$ on ${\cal C}$.}
 \label{fig:f5}
\end{figure*}
Assign a label $\{4,5,6\}-\{f'(u_1w_1),f'(u_{2t+1}w_{2t+1})\}$
to $(u_1,u_{2t+1})$ and
 a label in 
$\{4,5,6\}-\{f'(u_1w_1),f'(u_1u_{2t+1})\}$ to $(u_1,u_2)$.
Reassign
a label in 
$\{4,5,6\}-\{f'(u_{2t+1}w_{2t+1}),f'(u_1u_{2t+1})\}$ to $(u_{2t},u_{2t+1})$.
Then if $f'(u_4u_5)=4$, then let
$f'(u_4):=6$.

(3.1.2) Assume that $f'(u_1 w_1)= 3$ and $f'(u_{2t+1} w_{2t+1})\neq 0
 $.

Let $f':=f_5$.
We reassign labels for some vertices and edges as follows.
Let $f'(u_3):=6$, $f'(u_2u_3):=3$, and $f'(u_1u_3):=4$.
Assign a label in $\{5,6\}-\{f'(u_{2t+1}w_{2t+1})\}$ to
$(u_1, u_{2t+1})$ and
a label in $\{5,6\}-\{f'(u_1u_{2t+1})\}$
to $(u_1,u_2)$.
Reassign 
a label in
$\{4,5,6\}-\{f'(u_{2t+1}w_{2t+1}),$ 
$f'(u_1u_{2t+1})\}$ to $(u_{2t},u_{2t+1})$.
Then if $f'(u_4u_5)=4$, then let
$f'(u_4):=0$ and
$f'(u_3u_4):=2$.


(3.1.3) Assume that $f'(u_1 w_1)\neq 3$ and $f'(u_{2t+1} w_{2t+1})=0$.

Let $f':=f_5$.

Consider the case of $t=2$.
Let $f'(u_3u_5):=4$ and $f'(u_3):=6$.
Next we assign a label in $\{5,6\}-\{f'(u_1w_1)\}$ to
$(u_1, u_5)$
and  
 a label in
$\{4,5,6\}-\{f'(u_1u_5),f'(u_1w_1)\}$ to $(u_1,u_2)$.
Reassign a label in 
$\{5,6\}-\{f'(u_1u_5)\}$ to $(u_4,u_5)$.

Consider the case of $t\geq 4$.
Let $f'(u_{2t-1}u_{2t+1}):=5$, $f'(u_{2t-2}):=1$,
and $f'(u_{2t-1}):=0$.
Next we assign a label in $\{4,6\}-\{f'(u_1w_1)\}$ to
$(u_1, u_{2t+1})$
and  
 a label in
$\{4,5,6\}-\{f'(u_1u_{2t+1}),f'(u_1w_1)\}$ to $(u_1,u_2)$.
Reassign a label in 
$\{4,6\}-\{f'(u_1u_{2t+1})\}$ to $(u_{2t},u_{2t+1})$.
Then if $f'(u_{2t}u_{2t+1})=4$, then
let $f'(u_{2t}):=6$, and
if $f'(u_{2t}u_{2t+1})=6$, then
let  $f'(u_{2t}):=4$.


(3.1.4) Assume that $f'(u_1 w_1)=3$ and $f'(u_{2t+1} w_{2t+1})=0$.

If $t=2$,
let 
$f'(u_2):=2$,
$f'(u_3):=0$,
$f'(u_4):=6$,
$f'(u_1u_2):=6$,
$f'(u_2u_3):=5$,
$f'(u_1u_3):=4$,
$f'(u_3u_4):=2$,
 $f'(u_4u_5):=4$, 
$f'(u_3u_5)=6$,
and
$f'(u_1u_5):=5$.

Consider the case of $t\geq 4$.
Let $f':=f_5$.
We reassign labels for some vertices and edges similarly to
(3.1.2) and (3.1.3).
Namely, 
let $f'(u_1u_2):=5$,
$f'(u_3):=6$,
$f'(u_1u_3):=4$,
$f'(u_{2t-2}):=1$,
$f'(u_{2t-1}):=0$,
$f'(u_{2t}):=6$,
$f'(u_{2t}u_{2t+1}):=4$, $f'(u_{2t-1}u_{2t+1}):=5$, and
$f'(u_1u_{2t+1}):=6$.


(3.2) Assume that $t$ is odd; $t=2k+1~(k\geq 1)$ . 

(3.2.1) Assume that $f'(u_1 w_1)\neq 3$ and $f'(u_{2t+1} w_{2t+1})\neq
 0$.

Let $f'(u_2):=0$,
$f'(u_3):=5$,
$f'(u_4):=2$,
$f'(u_5):=6$,
$f'(u_6):=0$,
 $f'(u_7):=2$, 
$f'(u_2 u_3):=2$,
$f'(u_1u_3):=3$, 
$f'(u_3u_4):=0$,
 $f'(u_4u_5):=4$,
$f'(u_3u_5):=1$,
$f'(u_5u_6):=2$, 
$f'(u_6u_7):=6$, and
$f'(u_5 u_7):=0$.
For $i=2,3,\ldots,k$,
let 
$f'(u_{4i}):=0$,
  $f'(u_{4i+1}):=6$, 
$f'(u_{4i+2}):=0$, 
$f'(u_{4i+3}):=2$,
 $f'(u_{4i-1}u_{4i}):=5$,
 $f'(u_{4i}u_{4i+1}):=3$, 
$f'(u_{4i-1}u_{4i+1}):=4$,
 $f'(u_{4i+1}u_{4i+2}):=2$,
$f'(u_{4i+2}u_{4i+3}):=6$, and
$f'(u_{4i+1}u_{4i+3}):=0$.
Denote this labeling $f'$ by $f_6$ (see Fig.~\ref{fig:f6}).
\begin{figure*}[ht]
  \begin{center}
   \resizebox{3.6truein}{2.88truein}{\includegraphics{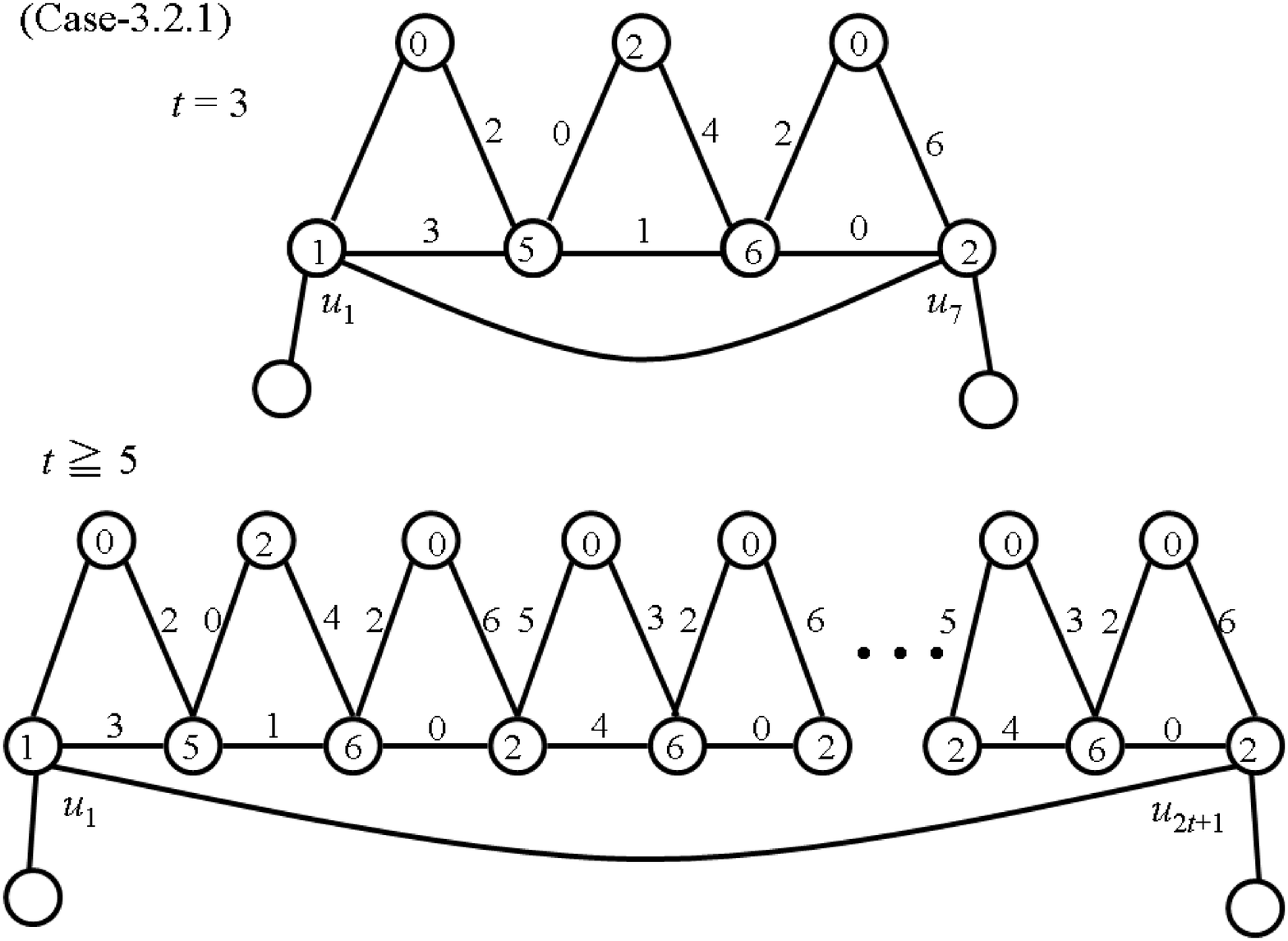}}
  \end{center}
 \caption{A labeling $f_6$ on ${\cal C}$.}
 \label{fig:f6}
\end{figure*}
We next assign a label $\{4,5,6\}-\{f'(u_1w_1),f'(u_{2t+1}w_{2t+1})\}$
to $(u_1,u_{2t+1})$ and
a label in
$\{4,5,6\}-\{f'(u_1w_1),f'(u_1u_{2t+1})\}$ to $(u_1,u_2)$.
Reassign
a label in 
$\{4,5,6\}-\{f'(u_{2t+1}w_{2t+1}),f'(u_1u_{2t+1})\}$ to
 $(u_{2t},u_{2t+1})$.
%

(3.2.2) Assume that $f'(u_1 w_1)= 3$ and $f'(u_{2t+1} w_{2t+1})\neq
 0$.

Let $f':=f_6$.
We reassign labels for some vertices and edges as follows.
Let $f'(u_1u_3):=6$ and
$f'( u_3):=4$.
Next we assign a label in $\{4,5\}-\{f'(u_{2t+1}w_{2t+1})\}$ to
$(u_1, u_{2t+1})$
and a label in
$\{4,5\}-\{f'(u_1u_{2t+1})\}$ to $(u_1,u_2)$.
Reassign a label in
$\{4,5,6\}-\{f'(u_1u_{2t+1}),f'(u_{2t+1}w_{2t+1})\}$ to
 $(u_{2t},u_{2t+1})$.


(3.2.3) Assume that $f'(u_1 w_1)\neq 3$ and $f'(u_{2t+1} w_{2t+1})=
 0$.

Let $f':=f_6$.

First consider the case of $t=3$.
Let 
$f'(u_4):=4$,
$f'(u_4u_5):=2$, 
$f'(u_5u_6):=3$, 
and
$f'(u_5u_7):=4$.
Next we assign a label in $\{5,6\}-\{f'(u_1w_1)\}$ to
$(u_1, u_7)$ 
and a label in
$\{4,5,6\}-\{f'(u_1w_1),f'(u_1u_7)\}$ to $(u_1,u_2)$.
Reassign
a label in
$\{5,6\}-\{f'(u_1u_7)\}$ to $(u_6,u_7)$.

Next consider the case of $t\geq 5$.
Let 
$f'(u_{2t-2}):=6$,
$f'(u_{2t-1}):=0$,
$f'(u_{2t-3}u_{2t-2}):=4$,
$f'(u_{2t-3}u_{2t-1}):=5$,
and
$f'(u_{2t-1}u_{2t+1}):=6$.
Next we assign a label in $\{4,5\}-\{f'(u_1w_1)\}$ to
$(u_1, u_{2t+1})$ 
and a label in
$\{4,5,6\}-\{f'(u_1w_1),f'(u_1u_{2t+1})\}$ to $(u_1,u_2)$.
Reassign
a label in
$\{4,5\}-\{f'(u_1u_{2t+1})\}$ to $(u_{2t},u_{2t+1})$.
Then if $f'(u_{2t}u_{2t+1})=4$ (resp., 5), then let
$f'(u_{2t}):=6$ (resp., 1) and $f'(u_{2t-1}u_{2t}):=2$ (resp., 4).

(3.2.4) Assume that $f'(u_1 w_1)= 3$ and $f'(u_{2t+1} w_{2t+1})=
 0$.

Let $f':=f_6$.
We reassign labels for some vertices and edges similarly to
(3.2.2) and (3.2.3).
Namely, if $t=3$,
 let 
$f'(u_1u_2):=4$,
$f'(u_1u_3):=6$,
$f'(u_3):=4$,
$f'(u_4):=5$,
$f'(u_3u_4):=1$,
$f'(u_4u_5):=2$,
$f'(u_3u_5):=0$,
$f'(u_5u_6):=3$,
$f'(u_6u_7):=6$,
$f'(u_5u_7):=4$, and
$f'(u_1u_7):=5$.
If $t\geq 5$, then
let $f'(u_1u_2):=5$,
$f'(u_1u_3):=6$,
$f'(u_3):=4$,
$f'(u_{2t-2}):=6$,
$f'(u_{2t-1}):=0$,
$f'(u_{2t}):=1$,
$f'(u_{2t-3}u_{2t-2}):=4$, 
$f'(u_{2t-3}u_{2t-1}):=5$,
$f'(u_{2t-1}u_{2t}):=4$,  
$f'(u_{2t}u_{2t+1}):=5$, 
 $f'(u_{2t-1}u_{2t+1}):=6$, and
$f'(u_1u_{2t+1}):=4$.

(Case-4) Let $f'(u_1):=2$ and $f'(u_{2t+1}):=3$.

(4.1) Assume that $t$ is even; $t=2k$. 

(4.1.1) Assume that $f'(u_1 w_1)\neq 4$ and $f'(u_{2t+1} w_{2t+1})\neq 1$.

Let 
$f'(u_2):=1$,
$f'(u_3):=6$,
$f'(u_4):=4$,
$f'(u_5):=3$,
$f'(u_2 u_3):=3$,
$f'(u_1u_3):=4$,
$f'(u_3u_4):=2$,
$f'(u_4u_5):=0$, and
$f'(u_3u_5):=1$.
For $i=2,3,\ldots,k$, let
$f'(u_{4i-2}):=2$,
$f'(u_{4i-1}):=4$,
$f'(u_{4i}):=5$,
$f'(u_{4i+1}):=3$,
$f'(u_{4i-3}u_{4i-2}):=5$,
 $f'(u_{4i-2}u_{4i-1}):=0$,
$f'(u_{4i-3}u_{4i-1}):=6$,
$f'(u_{4i-1}u_{4i}):=2$,
$f'(u_{4i}u_{4i+1}):=0$, and
$f'(u_{4i-1}u_{4i+1})$ $:=1$.
Denote this labeling $f'$ by $f_7$ (see Fig.~\ref{fig:f7}).
\begin{figure*}[ht]
  \begin{center}
   \resizebox{3.6truein}{2.88truein}{\includegraphics{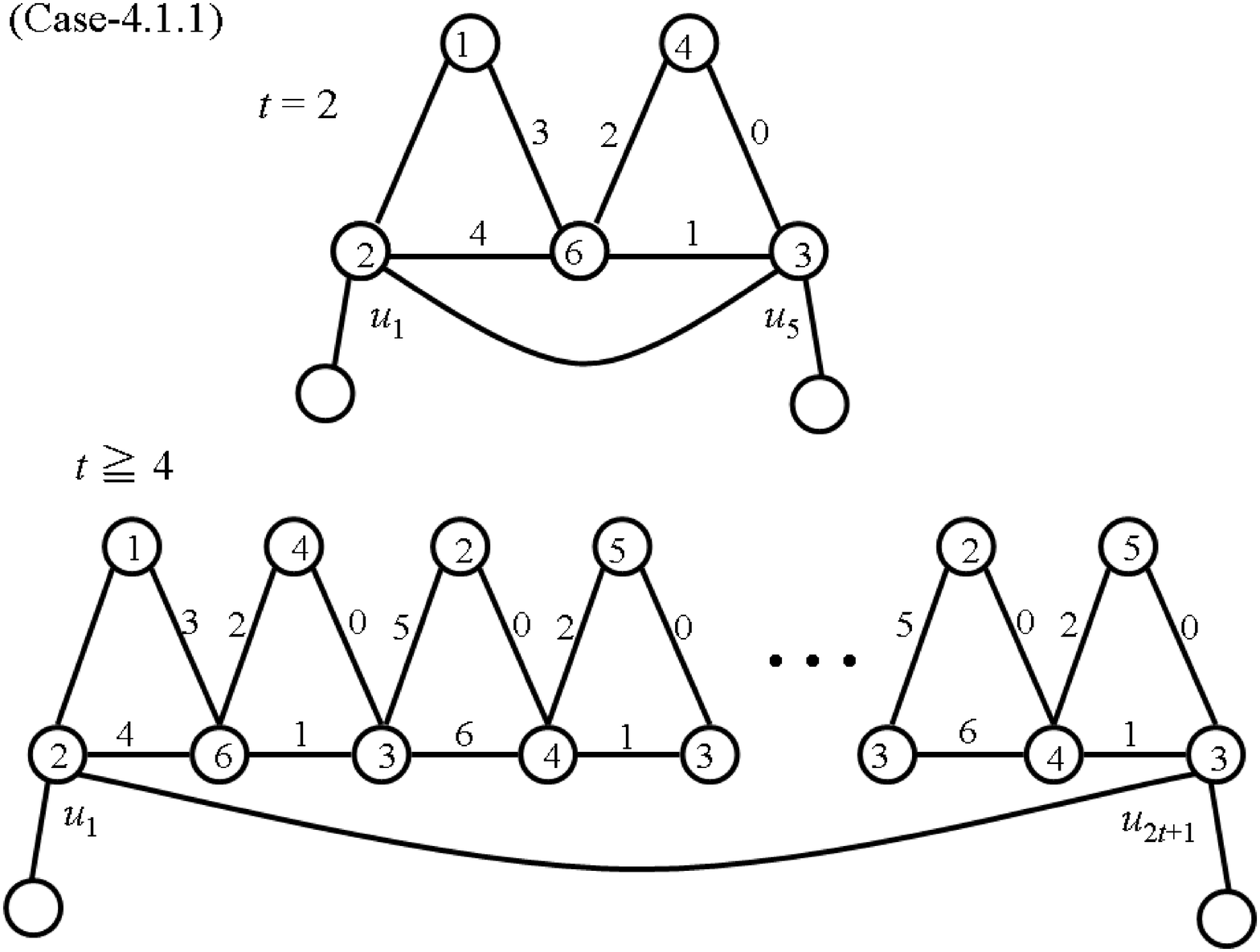}}
  \end{center}
 \caption{A labeling $f_7$ on ${\cal C}$.}
 \label{fig:f7}
\end{figure*}
Assign a label $\{0,5,6\}-\{f'(u_1w_1),f'(u_{2t+1}w_{2t+1})\}$
to $(u_1,u_{2t+1})$ and
 a label in 
$\{0,5,6\}-\{f'(u_1w_1),f'(u_1u_{2t+1})\}$ to $(u_1,u_2)$.
Reassign a label in 
$\{0,5,6\}-\{f'(u_{2t+1}w_{2t+1}),f'(u_1u_{2t+1})\}$ to $(u_{2t},u_{2t+1})$.
Then if $f'(u_1u_2)=0$, then let $f'(u_2):=5$, and
if $f'(u_{2t}u_{2t+1})\in \{5,6\}$, then let $f'(u_{2t}):=0$.

(4.1.2) Assume that $f'(u_1 w_1)= 4$ and $f'(u_{2t+1} w_{2t+1})\neq 1
 $.

Let $f':=f_7$.
We reassign a label for $(u_1, u_3)$ as
 $f'(u_1u_3):=0$.
Assign a label in $\{5,6\}-\{f'(u_{2t+1}w_{2t+1})\}$ to
$(u_1, u_{2t+1})$ and
a label in $\{5,6\}-\{f'(u_1u_{2t+1})\}$
to $(u_1,u_2)$.
Reassign 
a label in
$\{0,5,6\}-\{f'(u_{2t+1}w_{2t+1}),$ 
$f'(u_1u_{2t+1})\}$ to $(u_{2t},u_{2t+1})$.
Then if $f'(u_{2t}u_{2t+1}) \in \{5,6\}$, then let
$f'(u_{2t}):=0$.


(4.1.3) Assume that $f'(u_1 w_1)\neq 4$ and $f'(u_{2t+1} w_{2t+1})=1$.

Let $f':=f_7$.

Consider the case of $t=2$.
Let $f'(u_3u_5):=0$ and $f'(u_4):=0$.
Next we assign a label in $\{5,6\}-\{f'(u_1w_1)\}$ to
$(u_1, u_5)$ and
 a label in
$\{0,5,6\}-\{f'(u_1u_5),f'(u_1w_1)\}$ to $(u_1,u_2)$.
Reassign
 a label in 
$\{5,6\}-\{f'(u_1u_5)\}$ to $(u_4,u_5)$.
Then if $f'(u_1u_2)=0$, then let $f'(u_2):=5$.

Consider the case of $t\geq 4$.
Let $f'(u_{2t-1}u_{2t+1}):=5$, $f'(u_{2t-2}):=1$,
 $f'(u_{2t-1}):=0$, $f'(u_{2t-2}u_{2t-1}):=3$,
and $f'(u_{2t}):=4$.
Next we assign a label in $\{0,6\}-\{f'(u_1w_1)\}$ to
$(u_1, u_{2t+1})$
and
 a label in
$\{0,5,6\}-\{f'(u_1u_{2t+1}),f'(u_1w_1)\}$ to $(u_1,u_2)$.
Reassign
 a label in 
$\{0,6\}-\{f'(u_1u_{2t+1})\}$ to $(u_{2t},u_{2t+1})$.
Then if $f'(u_1u_2)=0$, then
let $f'(u_2):=5$.


(4.1.4) Assume that $f'(u_1 w_1)=4$ and $f'(u_{2t+1} w_{2t+1})=1$.

If $t=2$,
let 
$f'(u_2):=4$,
$f'(u_3):=0$,
$f'(u_4):=1$,
$f'(u_1u_2):=6$,
$f'(u_2u_3):=2$,
$f'(u_1u_3):=5$,
$f'(u_3u_4):=3$,
 $f'(u_4u_5):=5$, 
$f'(u_3u_5)=6$,
and
$f'(u_1u_5):=0$.

Consider the case of $t\geq 4$.
Let $f':=f_7$.
We reassign labels for some vertices and edges similarly to
(4.1.2) and (4.1.3).
Namely, 
let $f'(u_1u_2):=5$,
$f'(u_1u_3):=0$,
$f'(u_{2t-2}):=1$,
$f'(u_{2t-1}):=0$,
$f'(u_{2t}):=4$,
$f'(u_{2t-2}u_{2t-1}):=3$, 
$f'(u_{2t-1}u_{2t+1}):=5$, 
and
$f'(u_1u_{2t+1}):=6$.

\begin{figure*}[ht]
  \begin{center}
   \resizebox{3.6truein}{2.88truein}{\includegraphics{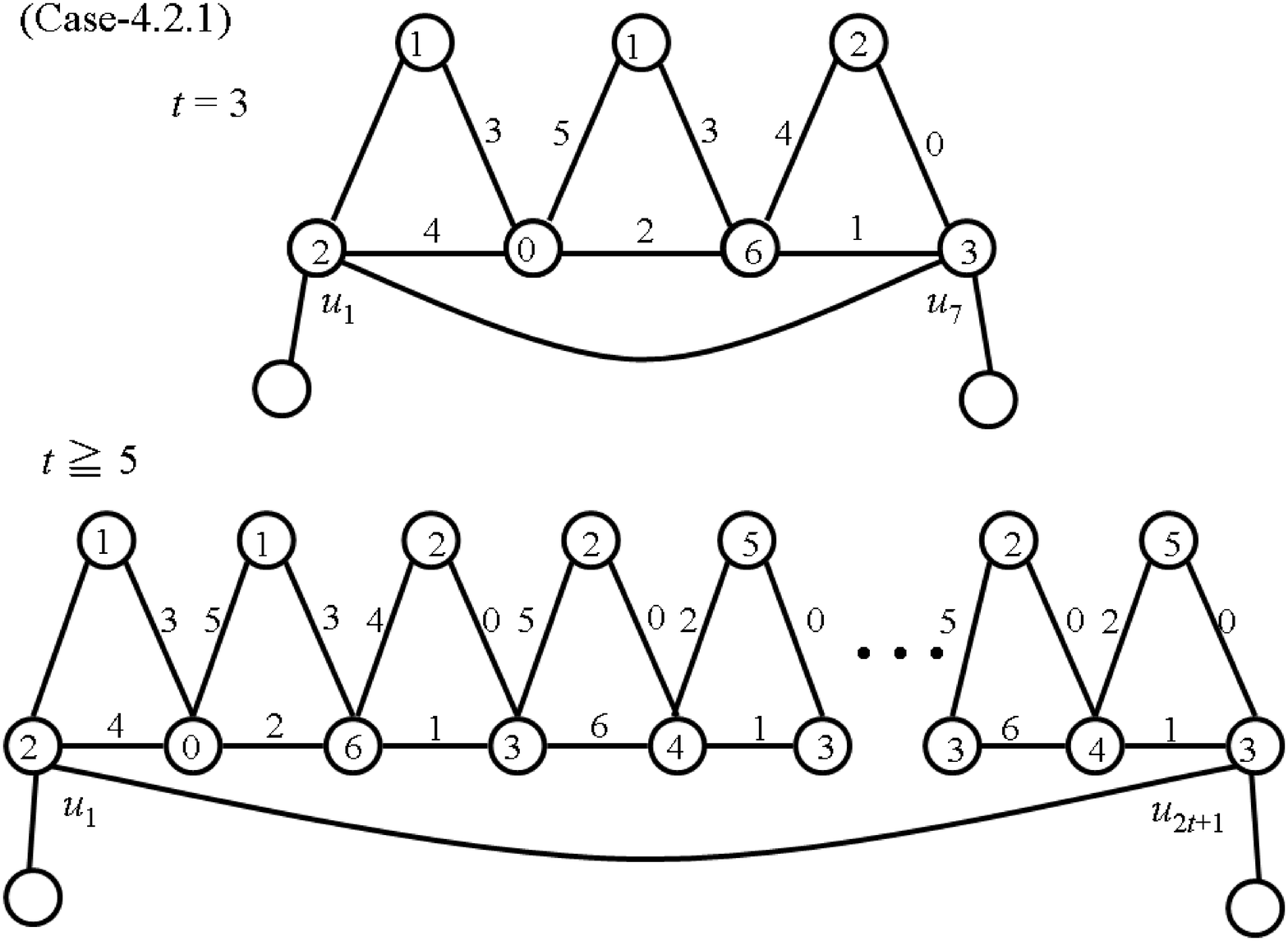}}
  \end{center}
 \caption{A labeling $f_8$ on ${\cal C}$.}
 \label{fig:f8}
\end{figure*}

(4.2) Assume that $t$ is odd; $t=2k+1~(k\geq 1)$ . 

(4.2.1) Assume that $f'(u_1 w_1)\neq 4$ and $f'(u_{2t+1} w_{2t+1})\neq
 1$.

Let $f'(u_2):=1$,
$f'(u_3):=0$,
$f'(u_4):=1$,
$f'(u_5):=6$,
$f'(u_6):=2$,
 $f'(u_7):=3$, 
$f'(u_2 u_3):=3$,
$f'(u_1u_3):=4$, 
$f'(u_3u_4):=5$,
 $f'(u_4u_5):=3$,
$f'(u_3u_5):=2$,
$f'(u_5u_6):=4$, 
$f'(u_6u_7):=0$, and
$f'(u_5 u_7):=1$.
For $i=2,3,\ldots,k$,
let 
$f'(u_{4i}):=2$,
  $f'(u_{4i+1}):=4$, 
$f'(u_{4i+2}):=5$, 
$f'(u_{4i+3}):=3$,
 $f'(u_{4i-1}u_{4i}):=5$,
 $f'(u_{4i}u_{4i+1}):=0$, 
$f'(u_{4i-1}u_{4i+1}):=6$,
 $f'(u_{4i+1}u_{4i+2}):=2$,
$f'(u_{4i+2}u_{4i+3}):=0$, and
$f'(u_{4i+1}u_{4i+3}):=1$.
Denote this labeling $f'$ by $f_8$ (see Fig.~\ref{fig:f8}).
We next assign a label $\{0,5,6\}-\{f'(u_1w_1),f'(u_{2t+1}w_{2t+1})\}$
to $(u_1,u_{2t+1})$ and
a label in
$\{0,5,6\}-\{f'(u_1w_1),f'(u_1u_{2t+1})\}$ to $(u_1,u_2)$.
Reassign 
a label in 
$\{0,5,6\}-\{f'(u_{2t+1}w_{2t+1}),f'(u_1u_{2t+1})\}$ to
 $(u_{2t},u_{2t+1})$.
Then if $f'(u_1u_2)=0$, then let $f'(u_2):=5$, and
if $f'(u_{2t}u_{2t+1})\in \{5,6\}$, then let $f'(u_{2t}):=0$.

%

(4.2.2) Assume that $f'(u_1 w_1)= 4$ and $f'(u_{2t+1} w_{2t+1})\neq
 1$.

Let $f':=f_8$.
We reassign a label for $(u_1, u_3)$ as
 $f'(u_1u_3):=6$.
Next we assign a label in $\{0,5\}-\{f'(u_{2t+1}w_{2t+1})\}$ to
$(u_1, u_{2t+1})$ and
 a label in
$\{0,5\}-\{f'(u_1u_{2t+1})\}$ to $(u_1,u_2)$.
Reassign a label in
$\{0,5,6\}-\{f'(u_1u_{2t+1}),f'(u_{2t+1}w_{2t+1})\}$ to
 $(u_{2t},u_{2t+1})$.
Then if $f'(u_1u_2)=0$, then let $f'(u_2):=5$, and
if $f'(u_{2t}u_{2t+1}) \in \{5,6\}$, then let $f'(u_{2t}):=0$.

(4.2.3) Assume that $f'(u_1 w_1)\neq 4$ and $f'(u_{2t+1} w_{2t+1})=
 1$.

Let $f':=f_8$.

First consider the case of $t=3$.
Let 
$f'(u_5u_7):=0$.
Next we assign a label in $\{5,6\}-\{f'(u_1w_1)\}$ to
$(u_1, u_7)$ and 
 a label in
$\{0,5,6\}-\{f'(u_1w_1),f'(u_1u_7)\}$ to $(u_1,u_2)$.
Reassign
a label in
$\{5,6\}-\{f'(u_1u_7)\}$ to $(u_6,u_7)$.
Then  if $f'(u_1u_2)=0$, then let $f'(u_2):=5$.

Next consider the case of $t\geq 5$.
Let 
$f'(u_{2t-2}):=0$,
$f'(u_{2t-1}):=1$,
$f'(u_{2t-2}u_{2t-1}):=4$,
$f'(u_{2t-1}u_{2t}):=3$,
and
$f'(u_{2t-1}u_{2t+1}):=5$.
Next we assign a label in $\{0,6\}-\{f'(u_1w_1)\}$ to
$(u_1, u_{2t+1})$ and
a label in
$\{0,5,6\}-\{f'(u_1w_1),f'(u_1u_{2t+1})\}$ to $(u_1,u_2)$.
Reassign 
a label in
$\{0,6\}-\{f'(u_1u_{2t+1})\}$ to $(u_{2t},u_{2t+1})$.
Then  if $f'(u_1u_2)=0$, then let $f'(u_2):=5$, and
if $f'(u_{2t}u_{2t+1})=6$, then let
$f'(u_{2t}):=0$.

(4.2.4) Assume that $f'(u_1 w_1)= 4$ and $f'(u_{2t+1} w_{2t+1})=
 1$.

Let $f':=f_8$.
We reassign labels for some vertices and edges similarly to
(4.2.2) and (4.2.3).
Namely, if $t=3$,
 let 
$f'(u_1u_2):=0$,
$f'(u_1u_3):=6$,
$f'(u_2):=5$,
$f'(u_6u_7):=6$,
$f'(u_5u_7):=0$, and
$f'(u_1u_7):=5$.
If $t\geq 5$, then
let $f'(u_1u_2):=5$,
$f'(u_1u_3):=6$,
$f'(u_{2t-2}):=0$,
$f'(u_{2t-1}):=1$,
$f'(u_{2t}):=0$,
$f'(u_{2t-2}u_{2t-1}):=4$, 
$f'(u_{2t-1}u_{2t}):=3$,  
$f'(u_{2t}u_{2t+1}):=6$, 
 $f'(u_{2t-1}u_{2t+1}):=5$, and
$f'(u_1u_{2t+1}):=0$.
\qed
\end{proof}


\end{document}